\documentclass[12pt, onecolumn, draftcls]{IEEEtran}
\newcommand{\iu}{\mathrm{i}}

\newtheorem{corollary}{Corollary}

\newtheorem{lemma}{Lemma}
\newtheorem{theorem}{Theorem}

\usepackage[ruled]{algorithm2e}

\usepackage[cmex10]{amsmath}
\allowdisplaybreaks
\DeclareMathOperator{\diag}{diag}
\DeclareMathOperator{\tr}{tr}

\usepackage{amssymb}
\newcommand{\naturals}{\mathbb{N}}
\newcommand{\reals}{\mathbb{R}}

\usepackage{color}
\usepackage{graphicx}
\usepackage{subfigure}
\usepackage{url}
\usepackage{tikz}

\begin{document}

\title{An Indirect Rate-Distortion Characterization for Semantic Sources: General Model and the Case of Gaussian Observation}
\author{Jiakun Liu, Shuo Shao, Wenyi Zhang, and H. Vincent Poor\thanks{J. Liu and W. Zhang are with Department of Electronic Engineering and Information Science, University of Science and Technology of China, Hefei, China (liujk@mail.ustc.edu.cn, wenyizha@ustc.edu.cn), S. Shao is with Department of Electronic Engineering, Shanghai Jiaotong University, Shanghai, China (shuoshao@sjtu.edu.cn), and H. Vincent Poor is with Department of Electrical Engineering, Princeton University, Princeton, NJ, USA (poor@princeton.edu). (Co-first authors: J. Liu and S. Shao; Corresponding authors: W. Zhang and H. V. Poor)

Preliminary results of this work have been presented in part at the IEEE International Symposium on Information Theory (ISIT), 2021 \cite{liu2021isit}.}}

\maketitle

\begin{abstract}
A new source model, which consists of an intrinsic state part and an extrinsic observation part, is proposed and its information-theoretic characterization, namely its rate-distortion function, is defined and analyzed. Such a source model is motivated by the recent surge of interest in the semantic aspect of information: the intrinsic state corresponds to the semantic feature of the source, which in general is not observable but can only be inferred from the extrinsic observation. There are two distortion measures, one between the intrinsic state and its reproduction, and the other between the extrinsic observation and its reproduction. Under a given code rate, the tradeoff between these two distortion measures is characterized by the rate-distortion function, which is solved via the indirect rate-distortion theory and is termed as the semantic rate-distortion function of the source. As an application of the general model and its analysis, the case of Gaussian extrinsic observation is studied, assuming a linear relationship between the intrinsic state and the extrinsic observation, under a quadratic distortion structure. The semantic rate-distortion function is shown to be the solution of a convex programming problem with respect to an error covariance matrix, and a reverse water-filling type of solution is provided when the model further satisfies a diagonalizability condition.
\end{abstract}


\section{Introduction}
\label{s:introduction}

A standard approach to describe an information source is to model a source as a stochastic process $\{X_i\}$, and when the stochastic process is memoryless, it suffices to model a source as a random variable\footnote{In this paper, random variables can be drawn from general alphabets, so random vectors are vector-valued random variables.} $X$ with a given probability distribution $p(x)$ \cite{shannon1948} \cite{cover2005}. In this paper, we study a new source model, which consists of an intrinsic state process and an extrinsic observation process. In the memoryless case, we can describe such a source model as a pair of random variables $(S, X)$, with a given joint probability distribution $p(s, x)$, defined over an appropriate product alphabet $\mathcal{S} \times \mathcal{X}$.

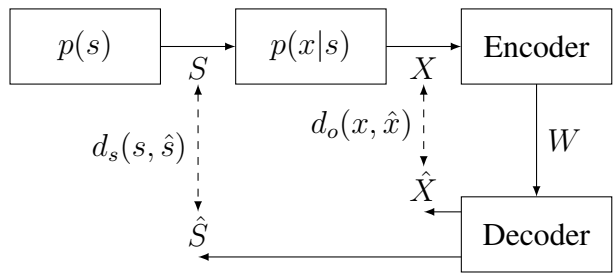
\begin{figure}[ht]
\centering
\begin{tikzpicture}
\draw (-1, -0.5) rectangle node {$p ( s )$} (1, 0.5);
\draw (2, -0.5) rectangle node {$p ( x | s )$} (4, 0.5);
\draw[-latex] (1, 0) -- node[below] {$S$} (2, 0);
\draw (5, -0.5) rectangle node {Encoder} (7, 0.5);
\draw[-latex] (4, 0) -- node[below] {$X$} (5, 0);
\draw (5, -3) rectangle node {Decoder} (7, -2);
\draw[-latex] (6, -0.5) -- node[right] {$W$} (6, -2);
\draw[-latex] (5, -2.2) -- (4.5, -2.2) node[above] {$\hat{X}$};
\draw[latex-latex, dashed] (4.5, -0.5) -- node[left] {$d_{o} ( x , \hat{x} )$} (4.5, -1.6);
\draw[-latex] (5, -2.8) -- (1.5, -2.8) node[above] {$\hat{S}$};
\draw[latex-latex, dashed] (1.5, -0.5) -- node[left] {$d_{s} ( s , \hat{s} )$} (1.5, -2.2);
\end{tikzpicture}
\caption{Illustration of a semantic source and its lossy compression.}
\label{f:model.schematic}
\end{figure}

In order to characterize the information-theoretic aspect of such a source, consider the problem of compressing the source $(S, X)$ so as to reproduce, in a lossy sense, a reproduction $(\hat{S}, \hat{X})$ over a reproduction product alphabet $\hat{\mathcal{S}} \times \hat{\mathcal{X}}$. Of course, a pair of distortion measures, $d_s: \mathcal{S} \times \hat{\mathcal{S}} \mapsto \mathbb{R}$ and $d_o: \mathcal{X} \times \hat{\mathcal{X}} \mapsto \mathbb{R}$, are introduced correspondingly. Here, the subscript $s$ stands for ``state'' and the subscript $o$ stands for ``observation''. A key point of the problem is that the compressor only has access to $X$, the extrinsic observation; --- while $S$, the intrinsic state, remains unrevealed. The situation is illustrated in Figure \ref{f:model.schematic}.

Our source model, termed as a semantic source in the sequel, is motivated by the recent surge of interest in the semantic aspect of information. In a number of applications that may benefit from taking into account the ``semantic'' feature of information, it is adequate to adopt a goal-oriented perspective; that is, the destination's interest in obtaining a piece of information is to accomplish a certain goal. Furthermore, it is customary to adopt an inference-theoretic problem formulation, which casts the accomplishment of the said goal as solving a statistical inference problem. The reproduction of the intrinsic state $S$ corresponds to the semantic inference part of the source, and the reproduction of the extrinsic observation $X$ corresponds to the conventional lossy compression part of the source.

We give two examples of the above consideration:
\begin{itemize}
    \item Systems that support MPEG Video Coding for Machines (VCM) are becoming popular in applications. In VCM, both the video itself and its features are reproduced: the video signal is for human vision, and the features are for machine vision tasks \cite{ma2019} \cite{duan2020} \cite{yang2021}. Treating the video as a semantic source, the video signal itself corresponds to its extrinsic observation, and the underlying features correspond to its intrinsic state, so as to embody the semantic aspect of the video. Usually the code rate required for reproducing features can be drastically lower than that required for reproducing the video signal itself. Intuitively, features typically have much smaller rate distortion functions and hence can be described with many fewer bits, compared with video signals. For instance, previous works have shown that neural network-based learning techniques can extract a very small amount of data from video signals to satisfy the need of action recognition, target classification and many other tasks \cite{yang2013} \cite{wu2021}.
    In contrast, traditional video coding schemes such as H.264/AVC/MPEG-4 and H.265/HEVC/MPEG-H Part 2 only target at reproducing the video signal with high fidelity, but may perform poorly for machine vision purposes \cite{liu2021}.
    
    \item In coding of speech signals, the semantic aspect is embodied as a sequence of text words, which, of course, can only be inferred from the speech signal itself. Treating the speech as a semantic source, the words correspond to its intrinsic state and the speech signal corresponds to its extrinsic observation. It is the usual case that both the words and the speech signal are desirable, because the words carry the meaning of speech, and the speech signal waveform may help us infer the stress and emotion of the speaker \cite{rabiner2007}, and may further help us accomplish tasks like speaker recognition and speaker verification \cite{furui1981}.
\end{itemize}

Our main contributions include:
\begin{itemize}
    \item We propose a theoretical framework based on rate distortion theory for characterizing semantic information.
    \item We define and derive a single-letter expression for the semantic rate distortion function.
    \item When the extrinsic observation is Gaussian and satisfies a linear relationship with the intrinsic state, we reduce the calculation of the semantic rate distortion function to a convex programming problem, which is tractable with standard scientific computing software. Furthermore, under a diagonalizability condition, we obtain a weighted reverse water-filling solution for the semantic rate distortion function.
\end{itemize}

We give a brief overview of related works in the remaining part of this section. Then we provide a formal mathematical description of the semantic source model and the corresponding semantic rate-distortion problem formulation in Section~\ref{s:model}, for which we establish the semantic rate-distortion function in general form in Section~\ref{s:general}. As an application of the general results, in Section~\ref{s:gaussian} we turn to a case study of Gaussian extrinsic observation, assuming a linear relationship between the intrinsic state and the extrinsic observation, under a quadratic distortion structure. Therein, we formulate a convex programming to solve the semantic rate-distortion function. When the Gaussian observation model further satisfies a diagonalizability condition, we develop a reverse water-filling type of solution in Section~\ref{u:water}. Finally we conclude this paper in Section~\ref{s:conclusion}.

\subsection{Related Works}
\label{ss:related-works}

The first formulation in Shannon's information theory is lossless source coding, wherein a sequence of symbols obeying a certain probabilistic law is represented as a bit string (i.e., a codeword) by an encoder, and the decoder reproduces, based upon the codeword, the original sequence of symbols, with success probability exactly one or asymptotically approaching one. Hence, the coding is solely determined by the probabilistic model of the source, and there is certainly no role of the semantic aspect of the source. This is also consistent with Shannon's remark in his landmark paper \cite{shannon1948}, saying ``these semantic aspects of communication are irrelevant to the engineering problem.''

In a broad sense, however, the lossy source coding formulation in Shannon's information theory, namely, the rate-distortion theory \cite{shannon1959}, has provided a means of studying the semantic aspects of a source. This is because the coding is not solely determined by the probabilistic model of a source, but is also affected by a distortion measure, which may be defined in a rather versatile way so as to capture the ``utility'' when the source is reproduced at the decoder.

Our present work goes one step further, by endowing a source with a state-observation structure and studying the rate distortion function of such a source model. This model captures the fact that the semantic aspects of a source are generally embedded as intrinsic features, and hence should be characterized by studying the reproduction of the intrinsic state, in addition to the reproduction of the extrinsic observation. Our treatment of semantic aspects of sources is also in line with the recently heightened interest in the development of 5G and beyond wireless systems \cite{popovski2019} \cite{kountouris2021} \cite{seo2021}, where for many applications the semantic aspects correspond to the accomplishment of certain inference goals. Hence, if we consider an information theoretic characterization of such a ``semantic'' source, the task of coding is to efficiently encode the extrinsic observation so that the decoder can infer both the intrinsic state and the extrinsic observation, subject to fidelity criteria on both, simultaneously. Our problem formulation and approach are closely related to two variants of the standard rate distortion theory, namely, indirect rate distortion function and rate distortion function under multiple distortion measures; see our discussion following Theorem~\ref{thm:general.rdf} in Section \ref{s:general}.


The inference-theoretic goal-oriented approach adopted in our problem formulation does not seek a task-independent universal definition of semantic information, which is outside the scope of the present paper; for some attempts in that regard, see, e.g., \cite{barhillel1953} \cite{floridi2004} \cite{bao2011} \cite{juba2011} for a few representative works that undertake drastically different approaches.

As related topics, the information bottleneck \cite{tishby1999} \cite{goldfeld2020} and the privacy funnel \cite{makhdoumi2014} \cite{shkel2021} are, in a certain sense, dual concepts, and both place constraints in terms of mutual information. The underlying idea of the information bottleneck is, in a broad sense, similar to ours. Specifically, there one generates a reproduction based upon the extrinsic observation, minimizing the mutual information between the extrinsic observation and the reproduction, while maintaining a level of mutual information between the intrinsic state and the reproduction. But for the information bottleneck problem formulation, there is neither explicit distortion measure, nor operational definition of lossy compression.

Task-based compression has been approached mainly from the perspective of quantizer design \cite{shlezinger2019}. It has been demonstrated that steering the design goal according to the task leads to performance benefits compared with conventional task-agnostic approach, a conclusion in line with what we advocate in our work. The perception-distortion tradeoff \cite{blau2019} imposes an additional constraint on the probability distribution of the reproduction. None of these related works proposes to decompose the information source into intrinsic and extrinsic parts as in our work, let alone investigate the joint behavior of them. In \cite{kipnis2021}, a similar intrinsic state-extrinsic observation model is studied, but the encoder is designed based on the marginal distribution of the extrinsic observation only.

\section{System Model and Problem Formulation}
\label{s:model}

As already outlined in the introduction, we model a memoryless semantic source as a pair of random variables $(S, X)$ that are correlated with joint probability distribution $p(s, x)$. The semantic aspect is embodied in the intrinsic state $S$, which is not observable but can only be inferred from the extrinsic observation $X$. In order to characterize the rate-distortion behavior of the semantic source, we consider a sequence of independent and identically distributed (i.i.d.) samples of $(S, X)$, denoted as $(S_i, X_i)_{i \in \naturals}$, and denote its length-$n$ block as $(S^n, X^n)$.


The i.i.d. source model is an idealistic scenario for our information-theoretic study. Real-world data generally exhibit sophisticated memory structures. A particularly interesting scenario is when the intrinsic state is a Markov chain, and the extrinsic observation obeys a hidden Markov model (HMM) \cite{rabiner1989}. Extensions of our approach for semantic source models with memory are left for future research.

The lossy compression of a semantic source has been illustrated in Figure \ref{f:model.schematic}. The encoder only has access to a length-$n$ block of the extrinsic observation sequence $X^n$, and the decoder has two tasks: reproducing the intrinsic state block as $\hat{S}^n$ under a state distortion measure $d_s$, and reproducing the extrinsic observation block as $\hat{X}^n$ under an observation distortion measure $d_o$. The encoder and the decoder are connected via a bit pipe in which the codeword $W$ of $nR$ bits is transferred from the encoder to the decoder, where $R$ is thus the code rate of the lossy compression system.

Below we provide a formal description of the lossy compression problem of a semantic source.

%

Let $d_{s} : \mathcal{S} \times \hat{\mathcal{S}} \to \reals_{+}$ and $d_{o} : \mathcal{X} \times \hat{\mathcal{X}} \to \reals_{+}$ be two given distortion measures, defined over the source product alphabet $\mathcal{S} \times \mathcal{X}$ and the reproduction product alphabet $\hat{\mathcal{S}} \times \hat{\mathcal{X}}$. The extended block-wise distortion measures are as follows:
\begin{align}
   & d_{s}(s^n,\hat{s}^n)=\frac{1}{n}\sum_{i=1}^{n}d_{s}(s_i,\hat{s}_i),\label{block error def 1}\\
   & d_{o}(x^n,\hat{x}^n)=\frac{1}{n}\sum_{i=1}^{n}d_{o}(x_i,\hat{x}_i)\label{block error def 2}.
\end{align}

We claim a tuple $(R,D_s,D_o)$ to be achievable, if for any $\epsilon>0$ and all sufficiently large $n$, there exist the following functions:
\begin{itemize}
    \item Encoding function $f: \mathcal{X}^n \mapsto \{1,2,\dots, 2^{\lfloor n(R+\epsilon)\rfloor }\}$ which generates the codeword $W$ as $W=f(X^n)$;
    
    \item State decoding function $g_s: \{1,2,\dots, 2^{\lfloor n(R+\epsilon)\rfloor} \} \mapsto \hat{\mathcal{S}}^n$, such that 
    \begin{align}\label{def:state-distortion}
        \mathbb{E}\left[d_{s}(S^n,\hat{S}^n)\right] \leq D_s+\epsilon,  
    \end{align}
    where $\hat{S}^n=g_s(f(X^n))$;
    
    \item Observation decoding function $g_o: \{1,2,\dots, 2^{\lfloor n(R+\epsilon)\rfloor} \} \mapsto \hat{\mathcal{X}}^n$, such that 
    \begin{align}\label{def:observation-distortion}
        \mathbb{E}\left[d_{o}(X^n,\hat{X}^n)\right] \leq D_o+\epsilon,  
    \end{align}
    where $\hat{X}^n=g_o(f(X^n))$.
\end{itemize}
It is clear that the state decoding function $g_s$ and the observation decoding function $g_o$ together constitute the decoder illustrated in Figure \ref{f:model.schematic}. 

Our goal is to characterize the region of all achievable $(R,D_s,D_o)$ tuples. Hence, we define the semantic rate distortion function as follows\footnote{This is the operational definition of a rate distortion function, which has been widely used (see, for example, \cite{cover2005} \cite{elgamal2011} \cite{yeung2008}).}:
\begin{align}
    R(D_s,D_o)=\inf \{R: (R,D_s,D_o)\;\text{is achievable} \}.
\end{align}
Clearly, characterizing the semantic rate distortion function $R(D_s,D_o)$ is equivalent to characterizing the achievable region of $(R,D_s,D_o)$.


We will also consider a variant of the distortion constraint; that is, the state distortion and the observation distortion are linearly combined to yield a single overall distortion. Hence, instead of \eqref{def:state-distortion} and \eqref{def:observation-distortion}, the decoding functions are required to satisfy the following weighted distortion constraint:
\begin{equation}\label{def:weighted-distortion}
    \mathbb{E} \left[
        w_{s} d_{s} ( S^{n} , \hat{S}^{n} )
        + w_{o} d_{o} ( X^{n} , \hat{X}^{n} )
    \right]
    \leq \bar{D} + \epsilon ,
\end{equation}
where $w_s$ and $w_o$ are non-negative weighting coefficients.

It is also natural to generalize the system model to include several intrinsic state variables each associated with a specified reproduction and a distortion. Such a semantic source is described by a tuple of random variables, $(S_0, S_1, \ldots, S_{k-1}, X)$, with joint probability distribution $p(s_0, s_1, \ldots, s_{k-1}, x)$ over $\mathcal{S}_0 \times \mathcal{S}_1 \times \ldots \times \mathcal{S}_{k-1} \times \mathcal{X}$, where each $S_j$ is an intrinsic state reflecting a certain semantic aspect of the source. The decoder now consists of an observation decoding function and $k$ state decoding functions, among which $g_{s, j}$ maps the codeword $W \in \{1, 2, \ldots, 2^{\lfloor n(R+\epsilon)\rfloor}\}$ into a reproduction sequence $\hat{S}_j^n$ to satisfy
\begin{align}
    \mathbb{E}\left[d_{s,j}(S_j^n, \hat{S}_j^n)\right] \leq D_{s,j} +\epsilon.
\end{align}
The notion of achievability can be defined in a similar fashion with respect to the tuple $(R$, $D_{s,0}$, $D_{s,1}$, $\ldots$, $D_{s,k-1}$, $D_o)$, and the semantic rate distortion function is consequently defined as
\begin{align}\label{def:rdf-multiple}
    R ( D_{s,0} , D_{s,1} , \cdots , D_{s, k - 1} , D_{o} )
    = \inf \{
        R :
        ( R , D_{s,0} , D_{s,1} , \cdots , D_{s,k - 1} , D_{o} )
        \; \text{is achievable}
    \} .
\end{align}

Examples of such semantic sources with multiple semantic aspects can be found in \cite{duan2020} \cite{liu2021}, which consider a hierarchy of image or video features, each feature associated with a quality metric.


\section{Semantic Rate Distortion Function}
\label{s:general}

In this section, we establish in the following theorem a single-letter characterization of the semantic rate distortion function $R(D_s, D_o)$ defined in Section \ref{s:model}.



\begin{theorem}
\label{thm:general.rdf}
For a given semantic source $(S, X)$ with $p(s, x)$ over $\mathcal{S} \times \mathcal{X}$, reproduction alphabet $\hat{\mathcal{S}} \times \hat{\mathcal{X}}$, and distortion measures $d_{s}$ and $d_{o}$, the semantic rate distortion function $R(D_s, D_o)$ is as follows:
\begin{align}
    R(D_s, D_o) &= \underset{p(\hat{s}, \hat{x}|x)} {\min} I(X; \hat{S}, \hat{X})\\
    \text{s.t.}\quad & \mathbb{E}\left[d_o(X, \hat{X})\right] \leq D_o,\\
         & \mathbb{E} \left[\hat{d}_{s} ( X , \hat{S} ) \right] \le D_{s},
\end{align}
where
\begin{align}
    \hat{d}_s(x, \hat{s}) = \mathbb{E}\left[d_s(S, \hat{s}) | x\right] = \sum_{s \in \mathcal{S}} p(s| x) d_s(s, \hat{s})
   \label{e:general.reducedd},
\end{align}
and $S, X, \hat{S}, \hat{X}$ constitute a Markov chain $S \leftrightarrow X \leftrightarrow (\hat{S}, \hat{X})$.
\end{theorem}

\textit{Proof:} See Appendix \ref{ proof of theorem general case}. $\Box$


Here we briefly discuss the basic idea of the proof of Theorem \ref{thm:general.rdf}. There are two main ingredients in the problem formulation: an indirect rate distortion problem which has been studied in \cite{dobrushin1962} \cite{wolf1970} \cite[Chap. 3, Sec. 5]{berger1971} \cite{witsenhausen1980}, and a rate distortion problem with several distortion constraints which has been studied in \cite[Sec. VII]{gamal1982} \cite[Prob. 10.19]{cover2005} \cite[Prob. 7.14]{csiszar2011}. A key is to recognize reproducing $\hat{S}$ as an indirect rate distortion problem, for which the state distortion between $S$ and $\hat{S}$ can be equivalently converted to a distortion between $X$ and $\hat{S}$. Indeed, the converted distortion is nothing but the conditional expectation of the original state distortion $d_s(S, \hat{s})$, over $p(s|x)$. This conversion hence circumvents the difficulty due to the absence of access to $S$ at the encoder. The detailed derivation, which is based on a unified treatment in \cite{witsenhausen1980}, is given in Appendix \ref{ proof of theorem general case}.

We note that the semantic rate distortion function can be non-trivial even for the special case where $S$ is a deterministic function of $X$, because from a lossy reproduction of $X$ it is generally impossible to reproduce $S$ in a lossless fashion. Specifically, suppose that $S = g (X)$. Then $\hat{d}_s(x, \hat{s})$ can be simplified into
\begin{align}
    \hat{d}_s(x, \hat{s}) = \sum_{s \in \mathcal{S}} p(s|x) d_s(s, \hat{s}) = d_s( g (x), \hat{s}).
\end{align}

Similar to standard rate distortion functions, a corollary of the semantic rate distortion function as given by Theorem~\ref{thm:general.rdf} is the following regarding monotonicity and convexity.
\begin{corollary}
\label{cor:monotonicity-convexity}
The semantic rate distortion function $R(D_s, D_o)$ in Theorem \ref{thm:general.rdf} has the following properties:
\begin{itemize}
    \item $R(D_s, D_o)$ is monotonically nonincreasing with $D_s$ and $D_o$.
    \item $R(D_s, D_o)$ is jointly convex with respect to $(D_s, D_o)$.
    \item The contour set $\left\{(D_s, D_o): R(D_s, D_o) \leq R \right\}$ is convex for any $R \geq 0$.
\end{itemize}
\end{corollary}

\textit{Proof:} The proof of the first two properties is exactly the same as that for standard rate distortion functions; see, e.g., \cite{cover2005}. The third property is then an immediate corollary of the second property. $\Box$

Corollary \ref{cor:monotonicity-convexity} implies a trade-off between the two distortions: for a given code rate, the smaller the state distortion, the larger the observation distortion, and vice versa.
Concrete numerical examples can be found in Section~IV, where Figures~2 and 4 plot the achievable regions of $( R , D_{s} , D_{o} )$ and their projections under different values of $R$, for two experimental setups, respectively. These plots demonstrate that for fixed $R$, the achievable $( D_{s} , D_{o} )$ pairs form a convex region, whose boundary exhibits a trade-off between $D_{s}$ and $D_{o}$.
Hence a sensible coding scheme of a semantic source should exhibit such behavior.


Now consider the weighted distortion constraint \eqref{def:weighted-distortion}. We have the following corollary.

\begin{corollary}
\label{cor:general.weighted}
For a given semantic source under the weighted distortion constraint \eqref{def:weighted-distortion}, the rate distortion function is as follows:
\begin{equation}
R(\bar{D}) = \min\left\{R(D_{s},D_{o}) | w_{s} D_{s} + w_{o} D_{o} \leq \bar{D}\right\}.
\end{equation}
\end{corollary}

\textit{Proof:} Given the semantic rate distortion function $R(D_s, D_o)$ in Theorem \ref{thm:general.rdf}, we have that any coding scheme that achieves $(R, \bar{D})$ should achieve a $(R, D_s, D_o)$ tuple for the semantic rate distortion problem under distortion constraints \eqref{def:state-distortion} and \eqref{def:observation-distortion}, for some $D_s$ and $D_o$ satisfying $w_{s} D_{s} + w_{o} D_{o} \leq \bar{D}$, and vice versa. $\Box$

We end this section with the semantic rate distortion function \eqref{def:rdf-multiple} for semantic sources with several intrinsic states, as given by the following corollary. Its proof is essentially identical to that of Theorem \ref{thm:general.rdf}.
\begin{corollary}
\label{cor:general.several}
For a semantic source $(S_0, S_1, \ldots, S_{k - 1}, X)$ with $p(s_0, s_1, \ldots, s_{k - 1}, x)$ over $\mathcal{S}_0 \times \mathcal{S}_1 \times \ldots \times \mathcal{S}_{k - 1} \times \mathcal{X}$, reproduction alphabet $\hat{\mathcal{S}}_0 \times \hat{\mathcal{S}}_1 \times \ldots \times \hat{\mathcal{S}}_{k - 1} \times \hat{\mathcal{X}}$, and distortion measures $\{d_{s_j}\}_{j = 0, 1, \ldots, k - 1}$ and $d_{o}$, the semantic rate distortion function $R(D_{s_0}, D_{s_1}, \ldots, D_{s_{k - 1}}, D_o)$ is as follows:
\begin{align}
    R(D_{s_0}, D_{s_1}, \ldots, D_{s_{k - 1}}, D_o) &= \underset{p(\hat{s}_0, \hat{s}_1, \ldots, \hat{s}_{k - 1}, \hat{x}|x)} {\min} I(X; \hat{S}_0, \hat{S}_1, \ldots, \hat{S}_{k - 1}, \hat{X})\\
    \text{s.t.}\quad & \mathbb{E}\left[d_o(X, \hat{X})\right] \leq D_o,\\
         & \mathbb{E} \left[\hat{d}_{s_j} ( X , \hat{S}_j ) \right] \le D_{s_j},\quad j = 0, 1, \ldots, k - 1,
\end{align}
where
\begin{align}
    \hat{d}_{s_j}(x, \hat{s}_j) = \mathbb{E}\left[d_{s_j}(S_j, \hat{s}_j) | x\right] = \sum_{s_j \in \mathcal{S}_j} p(s_j| x) d_{s_j}(s_j, \hat{s}_j),
\end{align}
and $S, X, \{\hat{S}_j\}_{j = 0, 1, \ldots, k - 1}, \hat{X}$ constitute a Markov chain $S \leftrightarrow X \leftrightarrow (\hat{S}_0, \hat{S}_1, \ldots, \hat{S}_{k - 1}, \hat{X})$.
\end{corollary}

\section{Gaussian Observation with Linear State-Observation Relationship}
\label{s:gaussian}


Theorem \ref{thm:general.rdf} establishes the general form of the semantic rate distortion function, which comes with an optimization problem, extending its counterpart in a standard rate distortion problem. In this section, we specialize the general result to a case where the extrinsic observation $X$ is Gaussian and the intrinsic state-extrinsic observation pair $(S, X)$ satisfies a linear relationship, under quadratic distortion measures.

The extrinsic observation $X$ obeys a multivariate Gaussian distribution $\mathcal{N}(0, \mathbf{K}_X)$,\footnote{We use $\mathbf{K}_{V}$ to denote the covariance matrix of a random column vector $V$.} where $\mathbf{K}_{X}$ is an $m \times m$ positive semi-definite matrix. The intrinsic state $S$ is given by
\begin{equation}
    S = \mathbf{H}X + Z,\label{eqn:linear-model}
\end{equation}
where $\mathbf{H}$ is an $l \times m$ matrix, and $Z$ is a random vector independent of $X$, with zero mean and covariance matrix $\mathbf{K}_Z$. Note that we neither restrict $Z$ to be Gaussian nor require $\mathbf{H}$ or $\mathbf{K}_Z$ to be full-rank. According to \eqref{eqn:linear-model}, the intrinsic state $S$ is a linear transformation of $X$, further disturbed by an independent component $Z$. This linear assumption holds for jointly Gaussian intrinsic state $S$ and extrinsic observation $X$, and can usually be extended to non-Gaussian models as well, either precisely or approximately, for example, when a linear estimator of $S$ conditioned upon $X$ can be obtained by traditional statistical methods, or by multilayer perceptron (MLP) neural networks alternatively \cite{xia2012}. On the other hand, note that the linear assumption no longer holds when one invokes nonlinear mappings, and deriving an analytical form of the corresponding semantic rate distortion function will generally be an extremely difficult task.

This model covers the special case where $(S, X)$ are jointly Gaussian. In fact, if $(S, X)$ are jointly Gaussian with zero mean and covariance matrix
\begin{equation}
    \begin{bmatrix}
        \mathbf{K}_S & \mathbf{K}_{SX} \\
        \mathbf{K}_{SX}^T & \mathbf{K}_X
    \end{bmatrix}, \label{e:gaussian.covmat} 
\end{equation}
we can represent $S$ according to
\begin{equation}
    S = \mathbf{K}_{SX} \mathbf{K}_X^{-1} X + Z, \label{e:gaussian.linear}
\end{equation}
where $Z \sim \mathcal{N}(0, \mathbf{K}_S - \mathbf{K}_{SX} \mathbf{K}_X^{-1} \mathbf{K}_{SX}^{T})$; that is, $\mathbf{H} = \mathbf{K}_{SX} \mathbf{K}_X^{-1}$ and $\mathbf{K}_Z = \mathbf{K}_S - \mathbf{K}_{SX} \mathbf{K}_X^{-1} \mathbf{K}_{SX}^{T}$.

We consider quadratic distortion measures, defined as
\begin{align}
        & d_{s} ( s , \hat{s} ) = \| s - \hat{s} \|_{2}^{2}=\tr( s - \hat{s} ) ( s - \hat{s} )^{T},\label{def:quadratic-state-distortion}\\
    & d_{o} ( x , \hat{x} ) = \| x - \hat{x} \|_{2}^{2}=\tr( x - \hat{x} ) ( x - \hat{x} ) ^{T}.\label{def:quadratic-observation-distortion}
\end{align}
Consequently, we have
\begin{align}
        &\mathbb{E} \left[ d_{s} ( S , \hat{S} ) \right] = \tr(\mathbf{K}_{S-\hat{S}}),\\
    &\mathbb{E} \left[ d_{o} ( X , \hat{X} ) \right] = \tr(\mathbf{K}_{X-\hat{X}}).
\end{align}

For the considered model \eqref{eqn:linear-model}, we can derive its semantic rate distortion function, given by the following theorem.


\begin{theorem}
\label{thm:gaussian.rdf}
The semantic rate distortion function for the semantic source with Gaussian extrinsic observation and linear state-observation relationship \eqref{eqn:linear-model}, under quadratic distortion measures \eqref{def:quadratic-state-distortion} and \eqref{def:quadratic-observation-distortion}, is given by:
\begin{align}
    R_{\mathcal{G}} ( D_{s} , D_{o} )
    & = \min_{\mathbf{\Delta} \in \mathcal{S}_m}  \frac{1}{2}
    \log \left( \frac{\det ( \mathbf{K}_{X} )}{\det ( \mathbf{\Delta} )} \right)    \label{e:gaussian.rdf} \\
\text{s.t.} \quad  & \mathbf{O} \prec \mathbf{\Delta}  \preceq \mathbf{K}_{X} ,
    \label{e:gaussian.deltale}
\\
 &    \tr ( \mathbf{H} \mathbf{\Delta} \mathbf{H}^{T} ) \le D_{s} - \tr ( \mathbf{K}_{Z} ) ,
    \label{e:gaussian.trds}
\\
&     \tr ( \mathbf{\Delta} ) \le D_{o} .
    \label{e:gaussian.trdx}
\end{align}
where $\mathcal{S}_m$ denotes the set of all $m \times m$ positive definite matrices. Note that here we use a subscript $\mathcal{G}$ to emphasize that the extrinsic observation is Gaussian.
\end{theorem}

\textit{Proof:} See Appendix \ref{u:gaussian.proof}. $\Box$

From \eqref{e:gaussian.trds}, when $Z$ is sufficiently strong so that $\tr (\mathbf{K}_Z) > D_s$, the optimization \eqref{e:gaussian.rdf} is no longer feasible and hence $R_\mathcal{G}(D_s, D_o) = \infty$. Otherwise, there is no further restriction on $\mathbf{K}_Z$. For example, even if $Z = 0$, i.e., the relationship between $S$ and $X$ is deterministic as $S = \mathbf{H}X$, the optimization problem in Theorem~\ref{thm:gaussian.rdf} is still non-trivial.

A simplified case arises when $\mathbf{H}$ is an orthogonal matrix satisfying $\mathbf{H}^T \mathbf{H} = \mathbf{I}$. In this case, \eqref{e:gaussian.trds} becomes
\begin{equation}
    \tr(\mathbf{H}\mathbf{\Delta} \mathbf{H}^T) = \tr (\mathbf{\Delta} \mathbf{H}^T \mathbf{H}) = \tr (\mathbf{\Delta}) \leq D_s - \tr (\mathbf{K}_Z),
\end{equation}
which can then be combined with \eqref{e:gaussian.trdx} leading to a single distortion constraint
\begin{equation}
    \tr(\mathbf{\Delta}) \leq \min\{D_o, D_s - \tr(\mathbf{K}_Z)\}.
\end{equation}

In Theorem \ref{thm:gaussian.rdf}, the matrix $\mathbf{\Delta}$ which we optimize corresponds to the mean squared error (MSE) of estimating $X$ based upon $\hat{X}$ at the decoder. The key to the proof of Theorem \ref{thm:gaussian.rdf} is to show that the semantic rate distortion function is achieved by a Gaussian reproduction. This is similar to situations in several Gaussian lossy compression problems, including the standard Gaussian rate distortion problem \cite{shannon1959} and the Gaussian quadratic CEO problem \cite{oohama1998}. Existing techniques based on the entropy power inequality (EPI), extremal inequalities, and Fisher information inequalities may also be interpreted as the optimality of Gaussian reproduction for the minimum mean squared error (MMSE) estimation under a given MSE constraint. In our analysis, we further need to accommodate with two MSE constraints, corresponding to the intrinsic state and the extrinsic observation, respectively.

Compared with the general form of semantic rate distortion function in Theorem \ref{thm:general.rdf}, Theorem \ref{thm:gaussian.rdf} involves only one matrix-valued optimization variable $\mathbf{\Delta}$, which, as remarked in the previous paragraph, is the MSE of estimating $X$ based upon $\hat{X}$ alone. In fact, the solution exhibits a Markov structure, i.e., $S \leftrightarrow X \leftrightarrow \hat{X} \leftrightarrow \hat{S}$. To help understand the optimality of the Markov chain solution, supposing that an alternative solution $(\hat{X}', \hat{S}')$ is given which does not satisfy the Markov structure,  consequently one can form an improved reproduction as $\hat{X} = \mathbb{E}(X | \hat{X}', \hat{S}')$, satisfying the Markov structure and achieving the same code rate $I(X; \hat{X} , \hat{S}')=I(X; \hat{X}',\hat{S}')$.

The Markov chain solution further suggests a ``two-stage'' coding interpretation which is in fact extensively adopted in practice: the decoder first generates a reproduction for $X$ as $\hat{X}$, and then uses that reproduction to further generate a reproduction for $S$ as $\hat{S}$. Similar to the standard Gaussian rate distortion problem, the optimal $\hat{X}$ can be constructed with the aid of a ``test channel'', for which $\hat{X}$ as the channel input is Gaussian and the additive Gaussian noise of the test channel has a covariance matrix $\Delta$, thereby producing $X$ as the desired channel output. To generate $\hat{S}$ based upon $\hat{X}$, it suffices to adopt a linear transform $\hat{S} = \mathbf{H} \hat{X}$. On the other hand, the Markov chain solution does not mean that the reproduction of $S$ is trivial, because the fidelity criterion on $X$ still needs to be adjusted according to $D_{s}$. The detailed arguments are given in the proof in Appendix \ref{u:gaussian.proof}.

An interesting property of the semantic rate distortion function derived in Theorem \ref{thm:gaussian.rdf} is that it is in fact an upper bound for all semantic sources with the same covariance structure under the quadratic distortion measure. This essentially indicates that a semantic source with Gaussian extrinsic observation is the hardest to describe, analogous to its counterpart in conventional source coding problems (see, e.g., \cite[Exercise 10.8]{cover2005}). Formally, we have the following corollary.


\begin{corollary}
    \label{cor:gaussian.largest_rdf}
For a semantic source $(S, X)$ with general probability density function, whose covariance matrix is given by \eqref{e:gaussian.covmat}, its semantic rate distortion function subject to quadratic distortion constraints \eqref{def:quadratic-state-distortion} and \eqref{def:quadratic-observation-distortion} satisfies $R ( D_{s} , D_{o} ) \le R_{\mathcal{G}}( D_{s} , D_{o} )$, where $R_{\mathcal{G}}( D_{s} , D_{o} )$ is the semantic rate distortion function given in Theorem \ref{thm:gaussian.rdf}, with $\mathbf{H} = \mathbf{K}_{SX} \mathbf{K}_X^{-1}$ and $\mathbf{K}_Z = \mathbf{K}_S - \mathbf{K}_{SX} \mathbf{K}_X^{-1} \mathbf{K}_{SX}^{T}$.
\end{corollary}

\textit{Proof:} See Appendix \ref{s:largest_rdf}. $\Box$

\subsection{Computation of the Semantic Rate Distortion Function}
\label{u:gaussian.computation}

We remark that the optimization problem in Theorem \ref{thm:gaussian.rdf} is convex, and hence can be numerically solved by software like CVX in an efficient and stable fashion. In this subsection we present some illustrative numerical examples.

\begin{figure}[htbp]
    \subfigure[Surface plot of $R_\mathcal{G} ( D_{s} , D_{o} )$]{
        \centering
        \includegraphics[width = 3.2in]{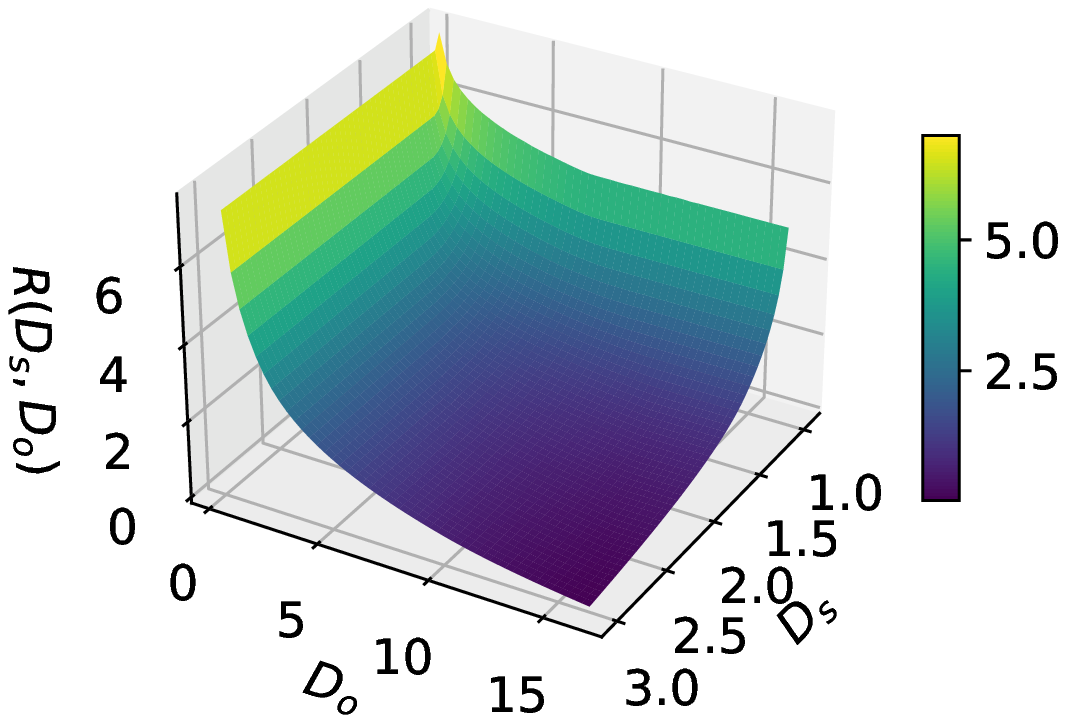}
        \label{f:gaussian.computation.toy3d}
    }
    \subfigure[Contour plot of $R_\mathcal{G} ( D_{s} , D_{o} )$]{
        \centering
        \includegraphics[width = 3.2in]{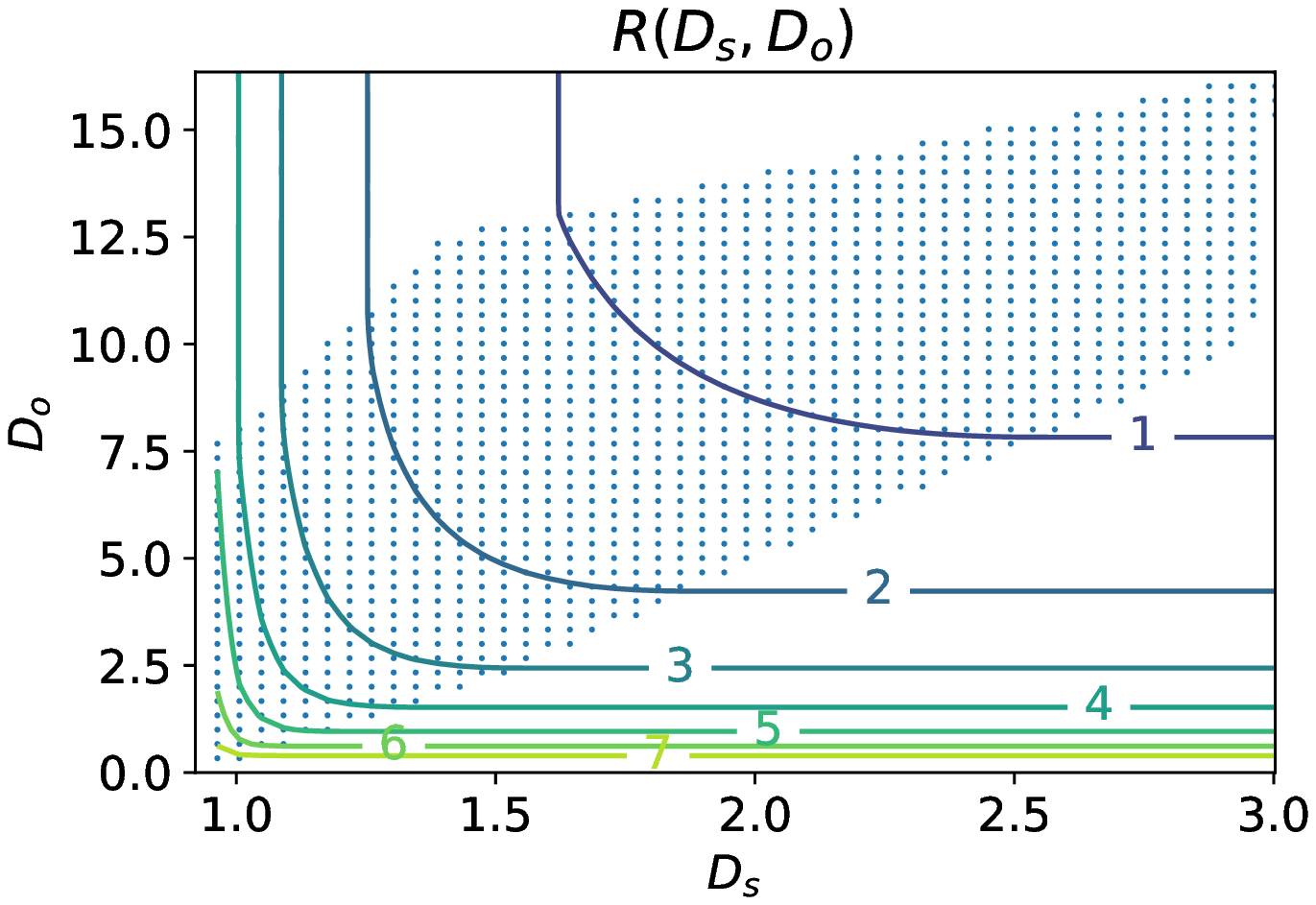}
        \label{f:gaussian.computation.toycontours}
    }
    \caption{Surface and contour plots of the semantic rate distortion function $R_\mathcal{G}(D_s, D_o)$ for the toy example.}
    \label{f:gaussian.computation.toy}
\end{figure}

Our first example is a small-scale toy model, given by
\begin{equation*}
    \mathbf{K}_{X} = \begin{bmatrix}
        11  & 0   & 0.5 \\
        0   & 3   & - 2 \\
        0.5 & - 2 & 2.35
    \end{bmatrix} , \;
    \mathbf{H} = \begin{bmatrix}
        0.0701   & 0.305   & 0.457 \\
        - 0.0305 & - 0.220 & 0.671
    \end{bmatrix} , \;
    \mathbf{K}_{Z} = \begin{bmatrix}
        0.701   & - 0.305 \\
        - 0.305 & 0.220
    \end{bmatrix}.
\end{equation*}
The resulting semantic rate distortion function is computed as displayed in Figure \ref{f:gaussian.computation.toy}. The dotted region in Figure \ref{f:gaussian.computation.toycontours} indicates that both constraints \eqref{e:gaussian.trds} and \eqref{e:gaussian.trdx} are active. The trade-off between the two distortions are clear: the smaller the state distortion, the larger the observation distortion, and vice versa.

\begin{figure}[htbp]
    \centering
    \includegraphics[width = 0.6\textwidth]{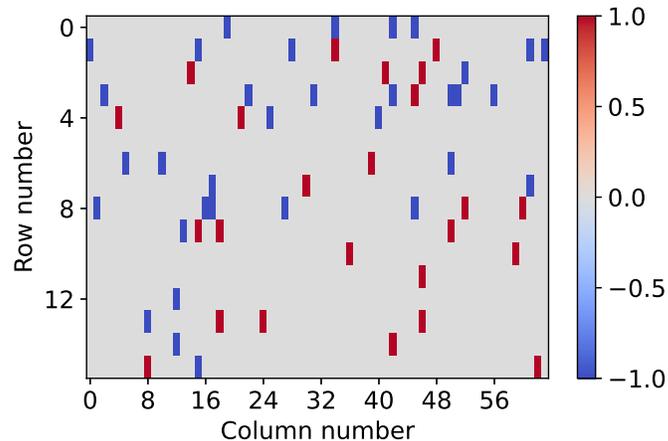}
    \caption{A $16 \times 64$ transformation matrix $\mathbf{H}$ shown as a two-dimensional grid. Elements are shown as cells with different colors corresponding to their values: blue for $-1$, red for $1$, and gray for $0$.}
    \label{f:gaussian.computation.transform}
\end{figure}

Our second example captures a sparse state-observation relationship, as follows. The extrinsic observation is a length-64 vector $X = [ X_{1}, \cdots, X_{64} ]^{T}$ consisting of i.i.d. $\mathcal{N} ( 0 , 2 )$ random variables. The transformation matrix $\mathbf{H}$ is a randomly masked $16 \times 64$ Rademacher matrix; that is, we first generate a Rademacher matrix whose elements are i.i.d. taking values $\{1, -1\}$ with equal probability $1/2$, and then independently reset these elements to zero with probability $0.95$. A realization of $\mathbf{H}$ is shown in 
Figure \ref{f:gaussian.computation.transform}. The noise vector $Z = [ Z_{1} , \cdots, Z_{16} ]^{T}$ consists of i.i.d. $\mathcal{N} ( 0 , 1 )$ random variables.

\begin{figure}[htbp]
	\subfigure[Surface plot of $R_\mathcal{G}(D_s, D_o)$]{
		\begin{minipage}[t]{0.5\linewidth}
			\centering
			\includegraphics[width=3.2in]{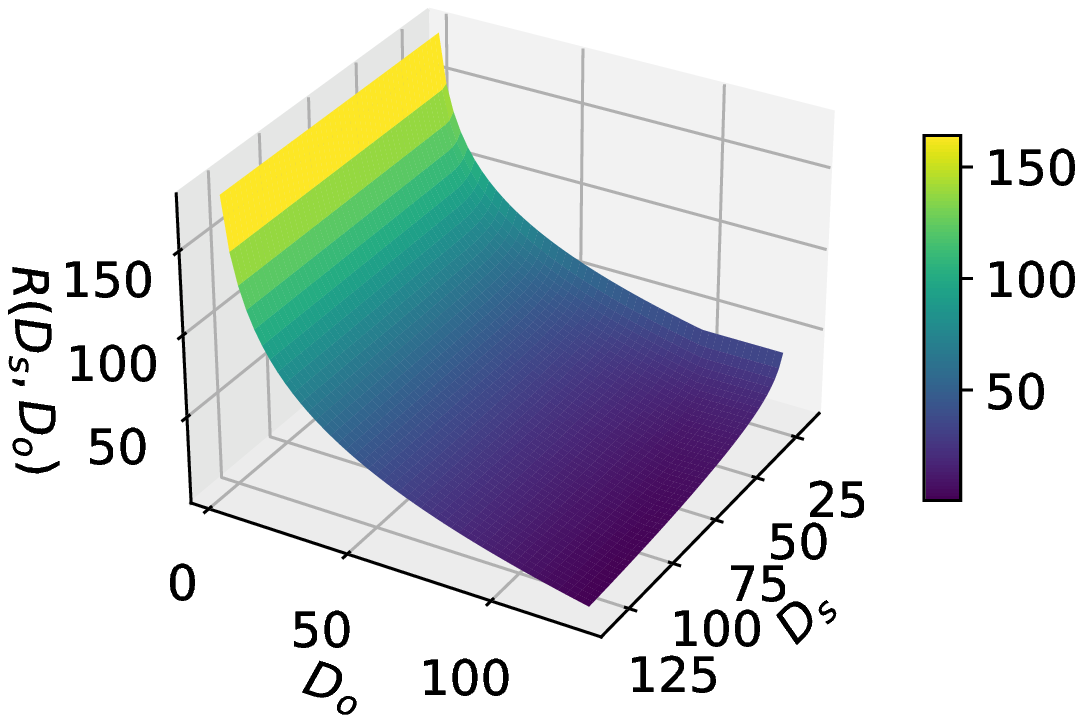}
	\end{minipage}
	\label{f:gaussian.computation.3d}}
	\subfigure[Contour plot of $R_\mathcal{G}(D_s, D_o)$]{
		\begin{minipage}[t]{0.5\linewidth}
			\centering
			\includegraphics[width=3.2in]{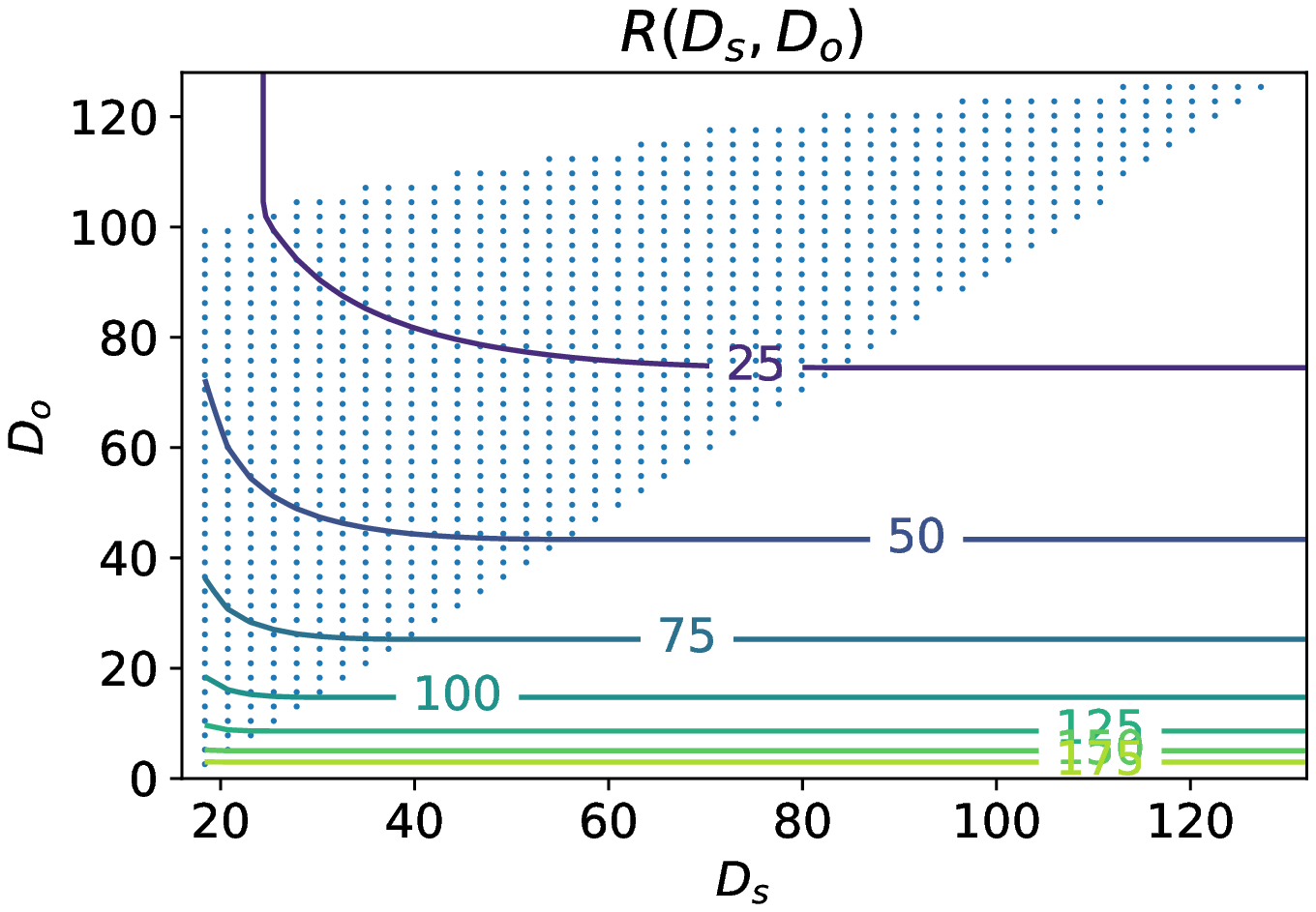}
	\end{minipage}
	\label{f:gaussian.computation.contours}}
	\caption{Surface and contour plots of the semantic rate distortion function $R_\mathcal{G}(D_s, D_o)$ for the example of a sparse state-observation relationship.}
	\label{f:gaussian.computation.sparse}
\end{figure}

We numerically solve the semantic rate distortion function according to Theorem~\ref{thm:gaussian.rdf}, and a typical surface of $R_\mathcal{G}(D_s, D_o)$ is illustrated in Figure \ref{f:gaussian.computation.3d}. More details of $R_\mathcal{G}(D_s, D_o)$ can be seen from the contour plot in Figure \ref{f:gaussian.computation.contours}, wherein the dotted region indicate that both constraints \eqref{e:gaussian.trds} and \eqref{e:gaussian.trdx} are active. From Figure \ref{f:gaussian.computation.contours}, it is evident that describing the extrinsic observation $X$ tends to be much more costly than describing the intrinsic state $S$: at the same code rate, the achieved $D_s$ is generally much lower than the achieved $D_o$.

\begin{figure}[htbp]
	\subfigure[$R_\mathcal{G}(D_s, D_o)$ versus $D_o$]{
		\begin{minipage}[t]{0.5\linewidth}
			\centering
			\includegraphics[width=3.3in]{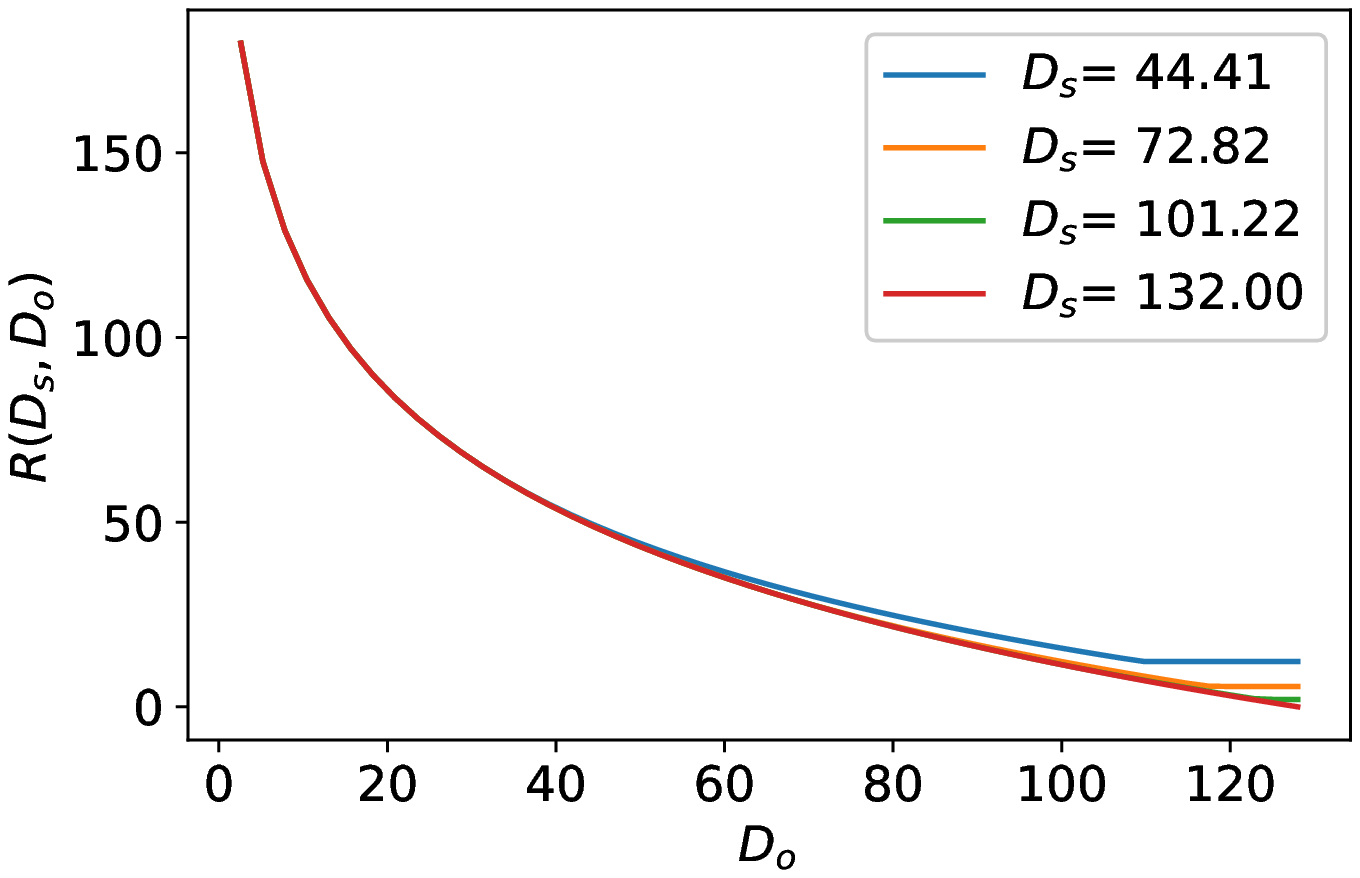}
	\end{minipage}
	\label{f:gaussian.computation.dobrate}}
	\subfigure[$R_\mathcal{G}(D_s, D_o)$ versus $D_s$]{
		\begin{minipage}[t]{0.5\linewidth}
			\centering
			\includegraphics[width=3.3in]{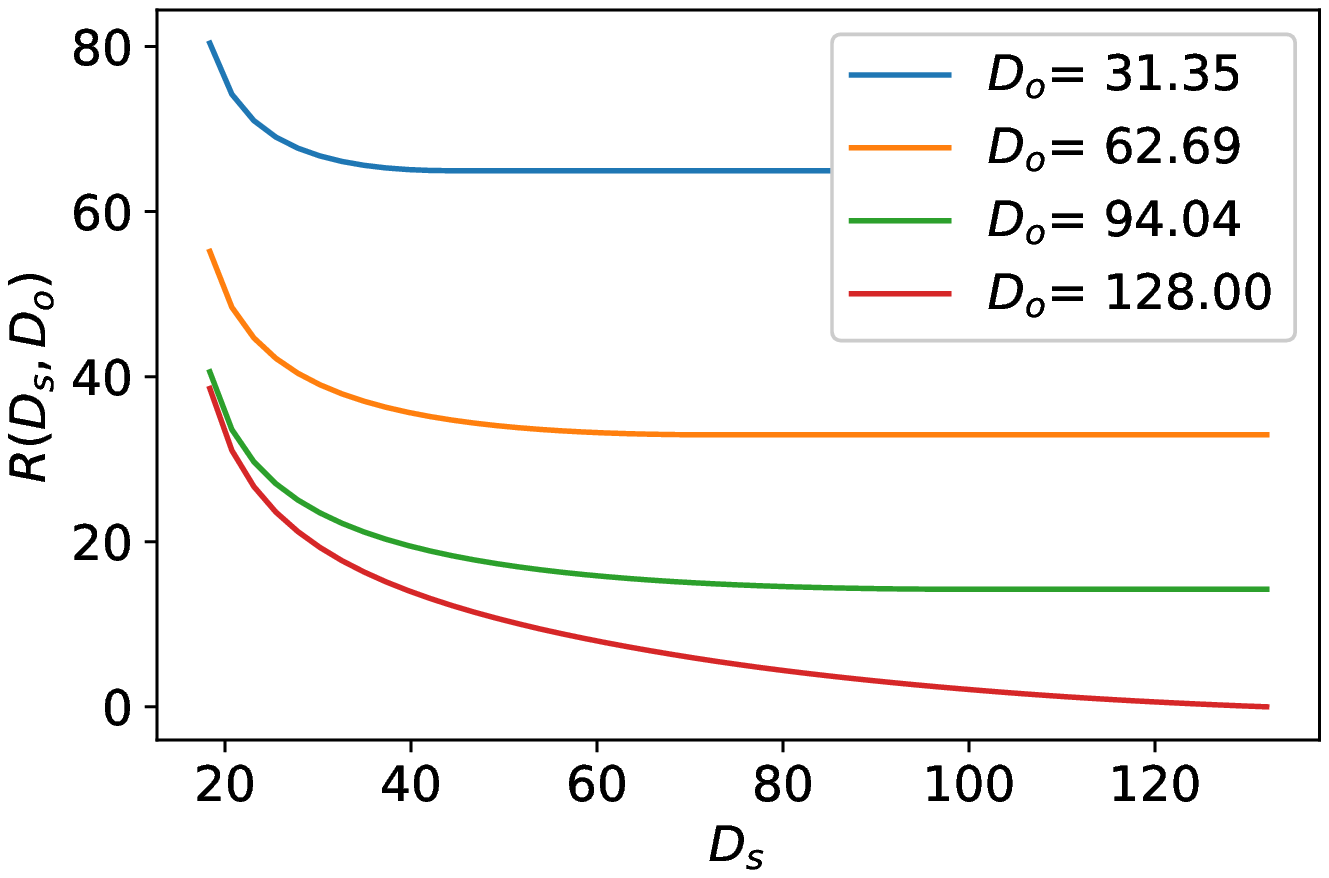}
	\end{minipage}
	\label{f:gaussian.computation.dstrate}}
	\caption{The semantic rate distortion function $R_\mathcal{G}(D_s, D_o)$ as a function of $D_o$ or $D_s$.}
\end{figure}

Another interesting fact regarding $R_\mathcal{G}(D_s, D_o)$ can be inferred from the dotted region in the contour plot Figure \ref{f:gaussian.computation.contours}, and is more clearly revealed by plotting the trends of $R_\mathcal{G}(D_s, D_o)$ as a function of $D_o$ (for fixed $D_s$) or $D_s$ (for fixed $D_o$), shown in Figures \ref{f:gaussian.computation.dobrate} and \ref{f:gaussian.computation.dstrate}, respectively. We find that, the code rate $R_\mathcal{G}(D_s, D_o)$ as a function of $D_o$ does not seem to be sensitive to the choice of $D_s$. This fact has an important consequence for designing lossy compression schemes for semantic sources: although several different codes may have similar performance in terms of reproducing the extrinsic observation, they can differ considerably in terms of reproducing the intrinsic state. A heuristic explanation is as follows: since $X$ is a high-dimensional vector, describing it along several different directions may lead to similar quadratic distortion performance; but since $S$ corresponds to a low-dimensional feature of $X$, its reproduction only favors the direction of describing $X$ that retains the feature of $S$ the best.

\subsection{Generalizations of Theorem~\ref{thm:gaussian.rdf}}
\label{u:gaussian.generalizations}

We can derive from Theorem \ref{thm:gaussian.rdf} several corollaries corresponding to the variants of the problem formulation in Section \ref{s:model}.

First, let us consider replacing the quadratic distortion measures by the positive semi-definite distortion constraints. Following the same arguments in the proof of Theorem \ref{thm:gaussian.rdf}, we again arrive at the optimality of Gaussian descriptions under positive semi-definite distortion constraints, and hence the following corollary characterizes the semantic rate distortion function.
\begin{corollary}
Consider the positive semi-definite distortion measures as 
\begin{align*}
        & d_{s} ( s , \hat{s} ) = ( s - \hat{s} ) ( s - \hat{s} )^{T},\\
    & d_{o} ( x , \hat{x} ) =( x - \hat{x} ) ( x - \hat{x} ) ^{T}.
\end{align*}
The semantic rate distortion function is given by \begin{align}
 R ( \mathbf{D}_{s} , \mathbf{D}_{o} )
    & = \min_{\mathbf{\Delta} \in \mathcal{S}_m}  \frac{1}{2}
    \log \left( \frac{\det ( \mathbf{K}_{X} )}{\det ( \mathbf{\Delta} )} \right)     \\
\text{s.t.}\quad  &  \mathbf{O} \prec \mathbf{\Delta}  \preceq \mathbf{K}_{X} ,
\\
 &     \mathbf{H} \mathbf{\Delta} \mathbf{H}^{T}  \preceq \mathbf{D}_{s} -  \mathbf{K}_{Z} ,
\\
&     \mathbf{\Delta} \preceq \mathbf{D}_{o}.
\end{align}
\end{corollary}

This is a semi-definite programming problem and can be readily solved by software.

Now consider the weighted distortion constraint, where the distortion measure is defined as a weighted sum of two individual distortion measures, i.e.
\begin{align}
    \bar{d} =    w_s d_{s} ( s , \hat{s} ) + w_o  d_{o} ( x , \hat{x} ) = w_s \| s - \hat{s} \|_{2}^{2}+ w_o  \| x - \hat{x} \|_{2}^{2}.
\end{align}

Applying Corollary \ref{cor:general.weighted}, we obtain the semantic rate distortion function in the following corollary.
\begin{corollary}
For the weighted distortion measure $\bar{d}$, the semantic rate distortion function $R(\bar{D})$ is given by
\begin{align}
     R ( \bar{D} )
    & = \min_{\mathbf{\Delta} \in \mathcal{S}_m}  \frac{1}{2}
    \log \left( \frac{\det ( \mathbf{K}_{X} )}{\det ( \mathbf{\Delta} )} \right)\\
\text{s.t.} \quad  & \mathbf{O}\prec \mathbf{\Delta}  \preceq \mathbf{K}_{X} ,
\\
& \tr ( ( w_{s} \mathbf{H}^{T} \mathbf{H} + w_{o} \mathbf{I}_{m} ) \mathbf{\Delta} )
    \leq \bar{D} - w_{s} \tr ( \mathbf{K}_{Z} ).
\end{align}
\end{corollary}

Finally, consider the case of $k$ intrinsic states. The extrinsic observation $X$ is still $\mathcal{N} ( 0 , \mathbf{K}_{X} )$. For each $j \in \{ 0 , 1 , \cdots , k - 1 \}$, the $j$-th intrinsic state is generated according to
\begin{equation*}
    S_{j} = \mathbf{H}_{j} X + Z_{j} ,
\end{equation*}
where $\mathbf{H}_{j}$ is an $l_{j} \times m$ matrix, and $Z_{j}$ is a random vector independent of $X$, with zero mean and covariance matrix $\mathbf{K}_{Z_{j}}$. We consider quadratic distortion measures, as
\begin{align}
    d_{s_{j}} ( s_j , \hat{s}_j ) &= \| s_j - \hat{s}_j \|_{2}^{2}, \quad j = 0, 1, \ldots, k - 1,\\
    d_{o}( x , \hat{x} ) &= \| x - \hat{x} \|_{2}^{2}.
\end{align}

The semantic rate distortion function is given by the following corollary.

\begin{corollary}
    \label{cor:gaussian.generalizations.multiple}
    For the semantic source with a Gaussian extrinsic observation and $k$ intrinsic states, the semantic rate distortion function under distortion measures $d_{s_{0}}$, $d_{s_{1}}$, $\cdots$, $d_{s_{k - 1}}$, $d_{o}$ is
    \begin{align*}
        R ( D_{s_{0}} , D_{s_{1}} , \cdots , D_{s_{k - 1}} , D_{o} )
        & = \min_{\mathbf{\Delta} \in \mathcal{S}_m}
        \frac{1}{2}
        \log \left( \frac{\det ( \mathbf{K}_{X} )}{\det ( \mathbf{\Delta} )} \right) \\
        \text{s.t.} \quad
        & \mathbf{O} \prec \mathbf{\Delta} \preceq \mathbf{K}_{X} , \\
        & \tr ( \mathbf{H}_{j} \mathbf{\Delta} \mathbf{H}_{j}^{T} ) \le D_{s_{j}} - \tr ( \mathbf{K}_{Z_{j}} ) , \quad
        j \in \{ 0 , 1 , \cdots , k - 1 \} , \\
        & \tr ( \mathbf{\Delta} ) \le D_{o} .
    \end{align*}
\end{corollary}

\section{Weighted Reverse Water-Filling}
\label{u:water}

Analogous to the standard Gaussian rate distortion problem wherein (after appropriate linear transformation) the solution can be interpreted as a reverse water-filling type of rate allocation, for the semantic rate distortion function in Theorem \ref{thm:gaussian.rdf}, under a diagonalizability condition, the solution can also be interpreted as reverse water-filling, but with appropriately weighted water levels.

For the model of Gaussian observation with linear state-observation relationship in Section~\ref{s:gaussian}, we further assume that the following diagonalizability condition is satisfied: there exists an unitary matrix $\mathbf{Q}$ such that
\begin{itemize}
    \item $\mathbf{Q}^{\dag} \mathbf{K}_{X} \mathbf{Q} = \diag ( \sigma_{1} , \sigma_{2} , \cdots , \sigma_{m} )$,
    \item $\mathbf{Q}^{\dag} \mathbf{H}^{T} \mathbf{H} \mathbf{Q} = \diag ( \alpha_{1} , \alpha_{2} , \cdots , \alpha_{m} )$
\end{itemize}
simultaneously hold. Here it loses no generality to order $\{\alpha_i\}_{i = 1}^m$ so that $\alpha_{1} \ge \alpha_{2} \ge \cdots \ge \alpha_{m}$. Denoting the rank of $\mathbf{H}^{T} \mathbf{H}$ as $q \le m$, then $\alpha_q > 0$ and $\alpha_{q + 1} = \cdots = \alpha_m = 0$.

\begin{lemma}
Under the diagonalizability condition, the resulting optimal $\mathbf{\Delta}$ takes the form
\begin{equation}
    \mathbf{\Delta} = \mathbf{Q} \diag ( \delta_{1} , \delta_{2} , \cdots , \delta_{m} ) \mathbf{Q}^\dag,
\end{equation}
and the semantic rate distortion function in Theorem~\ref{thm:gaussian.rdf} can be further written in terms of the following optimization problem:
\begin{align}
    R_\mathcal{G} ( D_{s} , D_{o} )
    & = \min_{\delta_{1} , \delta_{2} , \cdots , \delta_{m}}
    \frac{1}{2}
    \sum_{j = 1}^{m}
    \log \left( \frac{\sigma_{j}}{\delta_{j}} \right)
    \label{e:water.rate}
\\
    \text{s.t.} \quad &
    0 < \delta_{j} \le \sigma_{j} ,
    \quad \forall j \in \{ 1 , 2 , \cdots , m \} ,
    \label{e:water.delta}
\\
    & \sum_{j = 1}^{m} \alpha_{j} \delta_{j}
    \le D_{s} - \tr ( \mathbf{K}_{Z} ) ,
    \label{e:water.sumds}
\\
    & \sum_{j = 1}^{m} \delta_{j}
    \le D_{o} .
    \label{e:water.sumdx}
\end{align}
\end{lemma}

\textit{Proof:} See Appendix~\ref{s:waterderi}. $\Box$

In order to describe the weighted reverse water-filling solution, we first introduce the following curves.
\begin{itemize}
    \item Curve $C_s$:
    \begin{equation}
    C_{s} = \left\{ \left. \left(
        \sum_{j = 1}^{m}
        \alpha_{j}
        \min \left( \sigma_{j} , \frac{1}{\lambda} \right)
        + \tr ( \mathbf{K}_{Z} ) ,
        \sum_{j = 1}^{m}
        \min \left( \sigma_{j} , \frac{1}{\lambda} \right)
    \right) \right| \lambda > 0 \right\} ,
    \end{equation}
    which starts from $( \tr ( \mathbf{H} \mathbf{K}_{X} \mathbf{H}^{T} + \mathbf{K}_{Z} ) , \tr ( \mathbf{K}_{X} ) )$ and ends at $( \tr ( \mathbf{K}_{Z} ) , 0 )$.
    \item Curve $C_o$:
    \begin{equation}
    C_{o} = \left\{ \left. \left(
        \sum_{j = 1}^{q}
        \alpha_{j}
        \min \left( \sigma_{j} , \frac{1}{\mu \alpha_{j}} \right)
        + \tr ( \mathbf{K}_{Z} ) ,
        \sum_{j = 1}^{q}
        \min \left( \sigma_{j} , \frac{1}{\mu \alpha_{j}} \right)
        + \sum_{j = q + 1}^{m} \sigma_{j}
    \right) \right| \mu > 0 \right\} ,
    \end{equation}
    which starts from $( \tr ( \mathbf{H} \mathbf{K}_{X} \mathbf{H}^{T} + \mathbf{K}_{Z} ) , \tr ( \mathbf{K}_{X} ) )$ and ends at $( \tr ( \mathbf{K}_{Z} ) , \sum_{j = q + 1}^{m} \sigma_{j} )$. Here, $\sum_{j = 1 + 1}^{m} \sigma_{j}$ is interpreted as 0 if $\mathbf{H}^T \mathbf{H}$ is full-rank and thus $q = m$.
\end{itemize}

We then introduce the following partitioning of the $( D_{s} , D_{o} )$ plane, based upon the curves $C_s$ and $C_o$:
\begin{itemize}
\item $A_{0} = \{ ( D_{s} , D_{o} ) | D_{s} \ge \tr ( \mathbf{H} \mathbf{K}_{X} \mathbf{H}^{T} + \mathbf{K}_{Z} ) , D_{o} \ge \tr ( \mathbf{K}_{X} ) \}$;
\item $A_{1}$: on the right of the curve $C_{s}$, and between the two horizontal lines $D_{o} = 0$ and $D_{o} = \tr ( \mathbf{K}_{X} )$;
\item $A_{2}$: above the curve $C_{o}$, and between the two vertical lines $D_{s} = \tr ( \mathbf{K}_{Z} )$ and $D_{s} = \tr ( \mathbf{H} \mathbf{K}_{X} \mathbf{H}^{T} + \mathbf{K}_{Z} )$;
\item $A_{3}$: surrounded by the curves $C_{s}$ and $C_{o}$, and the vertical line $D_{s} = \tr ( \mathbf{K}_{Z} )$.
\end{itemize}
An example of the partitioning above is plotted in Figure \ref{f:water.cregions}.

The following theorem describes the weighted reverse water-filling solution.

\begin{theorem}
\label{thm.weighted-reverse-water-filling}
For the model of Gaussian observation with linear state-observation relationship in Section \ref{s:gaussian}, under the diagonalizability condition, the optimal $\mathbf{\Delta} = \mathbf{Q} \diag(\delta_1, \delta_2, \cdots \delta_m) \mathbf{Q}^\dag$ is given by
\begin{itemize}
    \item If $( D_{s} , D_{o} ) \in A_{0}$:
    \begin{equation}
    \delta_{j}^{*} = \sigma_{j} ,
    \quad \forall j \in \{ 1 , 2 , \cdots , m \} .
    \label{e:water.solution0}
    \end{equation}
    \item If $( D_{s} , D_{o} ) \in A_{1}$:
    \begin{equation}
    \delta_{j}^{*}
    = \min \left( \sigma_{j} , \frac{1}{\lambda} \right) ,
    \quad \forall j \in \{ 1 , 2 , \cdots , m \} ,
    \label{e:water.solution1}
    \end{equation}
    where $\lambda$ is chosen to satisfy $\sum_{j = 1}^{m} \delta_{j}^{*} = D_{o}$.
    \item If $( D_{s} , D_{o} ) \in A_{2}$:
    \begin{align}
    \delta_{j}^{*}
    = \begin{cases}
        \min \left( \sigma_{j} , \dfrac{1}{\mu \alpha_{j}} \right) , & \alpha_{j} > 0 \\
        \sigma_{j} , & \alpha_{j} = 0
    \end{cases} , \quad
    \forall j \in \{ 1 , 2 , \cdots , q \} ,
    \label{e:water.solution2}
    \end{align}
    where $\mu$ is chosen to satisfy $\sum_{j = 1}^{q} \alpha_{j} \delta_{j}^{*} = D_{s} - \tr ( \mathbf{K}_{Z} )$.
    \item If $( D_{s} , D_{o} ) \in A_{3}$:
    \begin{equation}
    \delta_{j}^{*}
    = \min \left( \sigma_{j} , \frac{1}{\lambda + \mu \alpha_{j}} \right) ,
    \quad \forall j \in \{ 1 , 2 , \cdots , m \} ,
    \label{e:water.solution3}
    \end{equation}
    where $\lambda$, $\mu$ are chosen to satisfy $\sum_{j = 1}^{m} \delta_{j}^{*} = D_{o}$ and $\sum_{j = 1}^{q} \alpha_{j} \delta_{j}^{*} = D_{s} - \tr ( \mathbf{K}_{Z} )$.
\end{itemize}
\end{theorem}

\textit{Proof:} See Appendix \ref{s:waterderi}. $\Box$

The partitioning $\{A_{0}, A_{1}, A_{2}, A_{3}\}$ is closely related to activity of the constraints \eqref{e:water.sumds} and \eqref{e:water.sumdx}, as summarized in Table~\ref{t:water.activity}. In $A_0$, both constraints are inactive, and hence the optimization is unconstrained yielding the trivial solution \eqref{e:water.solution0}. In $A_1$, only the observation distortion constraint is active, and the solution  \eqref{e:water.solution1} is a standard reverse water-filling with water level $1 / \lambda$. In $A_2$, only the state distortion is active, and the solution \eqref{e:water.solution2} essentially makes the weighted eigenvalues $\alpha_{1} \delta_{1}$, $\alpha_{2} \delta_{2}$, $\cdots$, $\alpha_{m} \delta_{m}$ fulfill a reverse water-filling structure, with water level $1 / \mu$. Alternatively, we may view the term $1 / (\mu \alpha_{j})$ in \eqref{e:water.solution2} as a water level with weight $1 / \alpha_{j}$. In $A_3$, both constraints are active, and the solution \eqref{e:water.solution3} also fulfills a reverse water-filling structure with unequal water levels.

\begin{table}
\centering
\caption{Activity of constraints \eqref{e:water.sumds} and \eqref{e:water.sumdx} in $A_{0}$, $A_{1}$, $A_{2}$ and $A_{3}$.}
\label{t:water.activity}
\begin{tabular}{c|cc}
    & \eqref{e:water.sumds} active & \eqref{e:water.sumds} inactive \\
    \hline
    \eqref{e:water.sumdx} inactive & $A_{2}$ & $A_{0}$ \\
    \eqref{e:water.sumdx} active & $A_{3}$ & $A_{1}$
\end{tabular}
\end{table}

\subsection{Case Study: Circulant $\mathbf{K}_X$ and $\mathbf{H}$ and Weighted Reverse Water-filling in Frequency Domain}

A case of special interest is where $\mathbf{K}_X$ and $\mathbf{H}$ are both circulant matrices \cite{gray2009}. As the dimension of $X$ grows large, this models the scenario where $X$ is a circularly stationary Gaussian process,\footnote{If we remove the circulant restriction and consider a stationary Gaussian process, then we encounter a Toeplitz $\mathbf{K}_X$, for which our solution still approximately applies; see, e.g., \cite{gray2009}.} and $S$ is obtained via passing $X$ through a time-invariant linear filter whose response is given by the first row of $\mathbf{H}$. For a circulant matrix, the corresponding unitary matrix $\mathbf{Q}$ is the well known discrete Fourier transform (DFT) matrix, and its eigenvalues are the DFT of the first row of the matrix. Hence the weighted reverse water-filling may be interpreted as exercised in the frequency domain, similar to its counterpart for the standard rate distortion function of stationary Gaussian processes.

\begin{figure}[htbp]
    \centering
    \begin{minipage}[t]{0.48\textwidth}
        \centering
        \includegraphics[width = 3.2in]{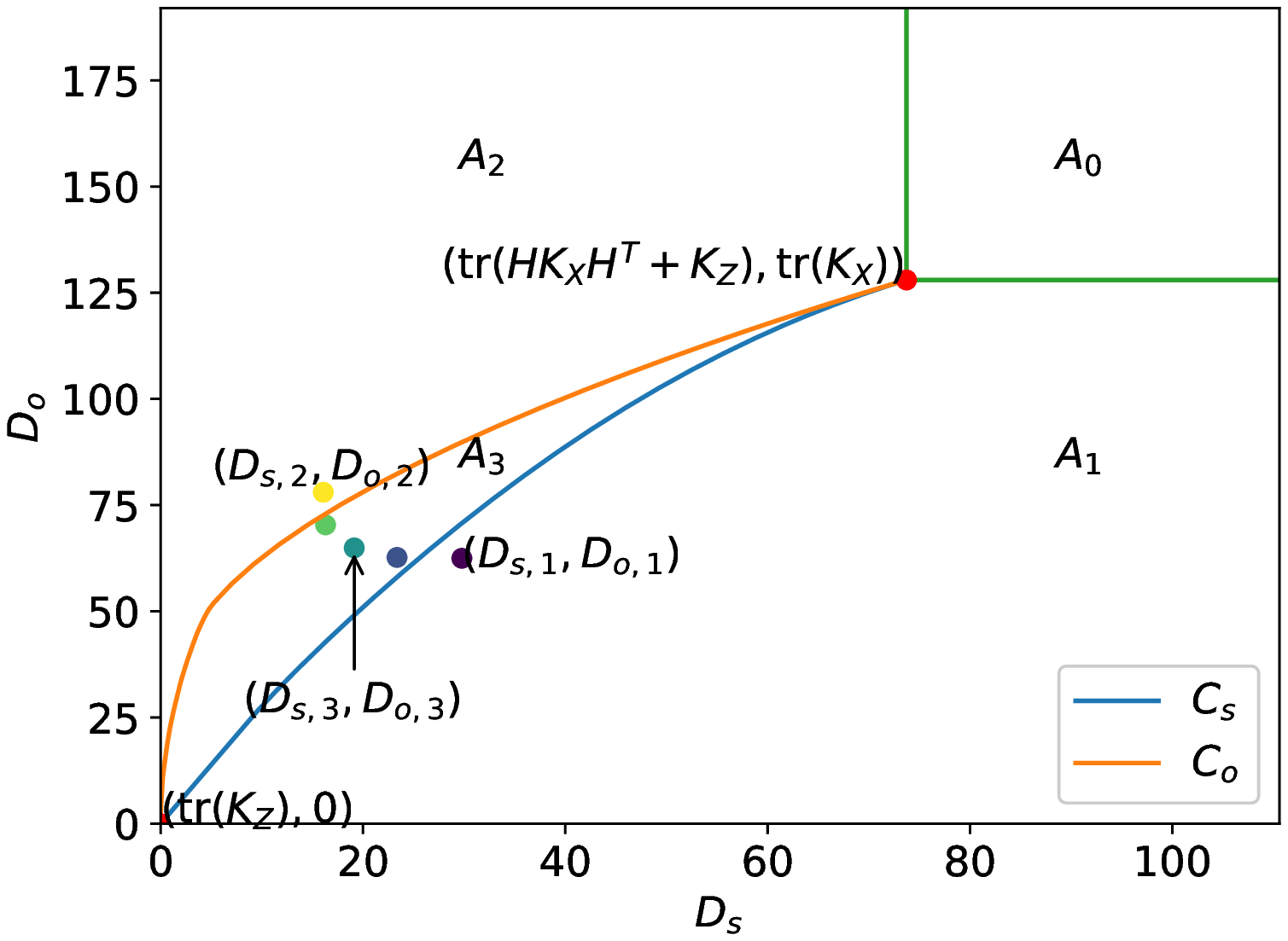}
        \caption{The $( D_{s} , D_{o} )$ plane is divided into four regions $A_{0}$, $A_{1}$, $A_{2}$, $A_{3}$, which determine the form of the optimal $\mathbf{\Delta}$. Five points on the contour $R_\mathcal{G} ( D_{s} , D_{o} ) = 50$ are marked with colors varying from purple to yellow.}
        \label{f:water.cregions}
    \end{minipage}
    \quad
    \begin{minipage}[t]{0.48\textwidth}
        \centering
        \includegraphics[width = 3.2in]{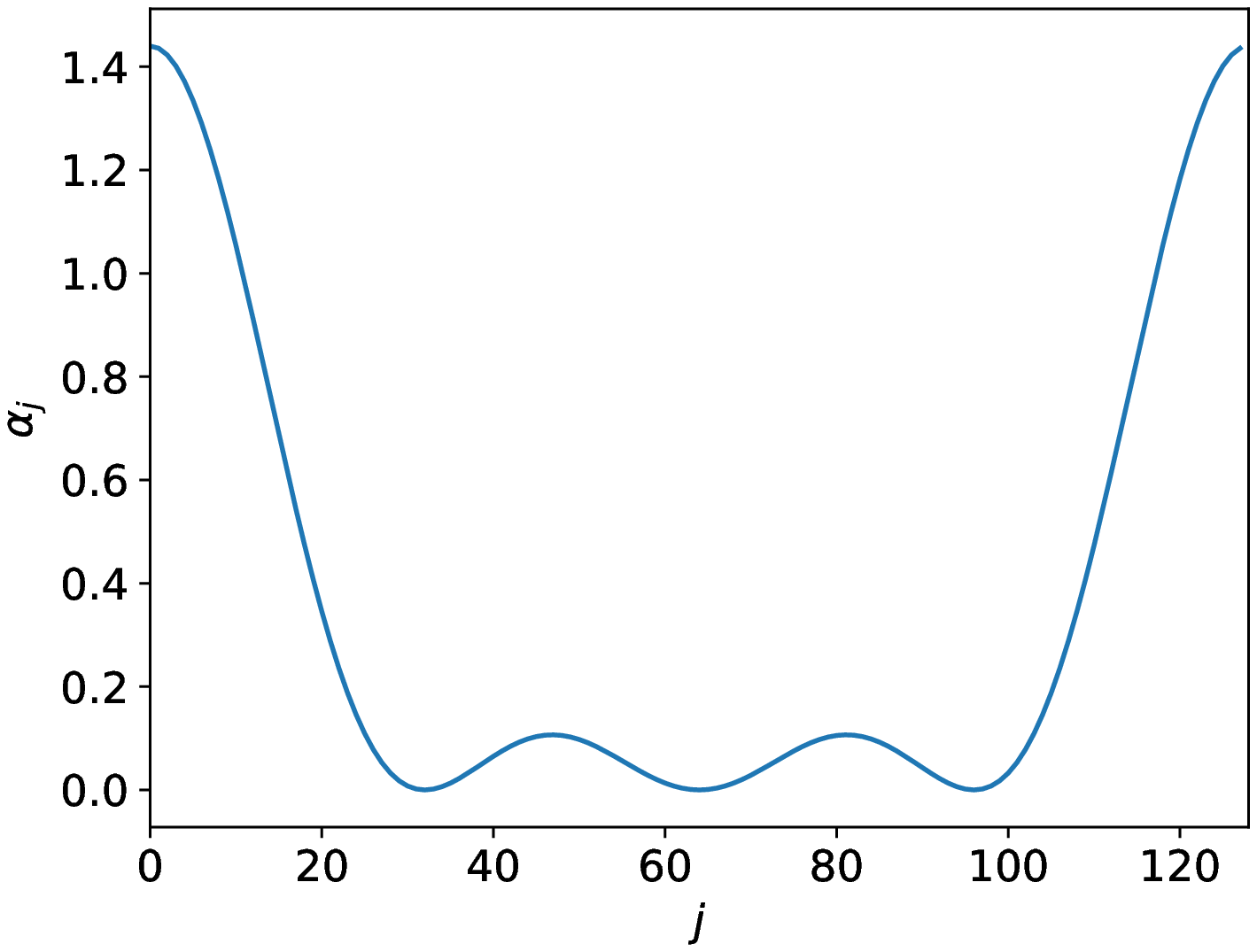}
        \caption{Diagonal elements $\alpha_{0}$, $\alpha_{1}$, $\cdots$, $\alpha_{127}$ of $\mathbf{Q}^{\dag} \mathbf{H}^{T} \mathbf{H} \mathbf{Q}$.}
        \label{f:water.alpha}
    \end{minipage}
\end{figure}

\begin{figure}[htbp]
    \centering
    \includegraphics[width = 4.8in]{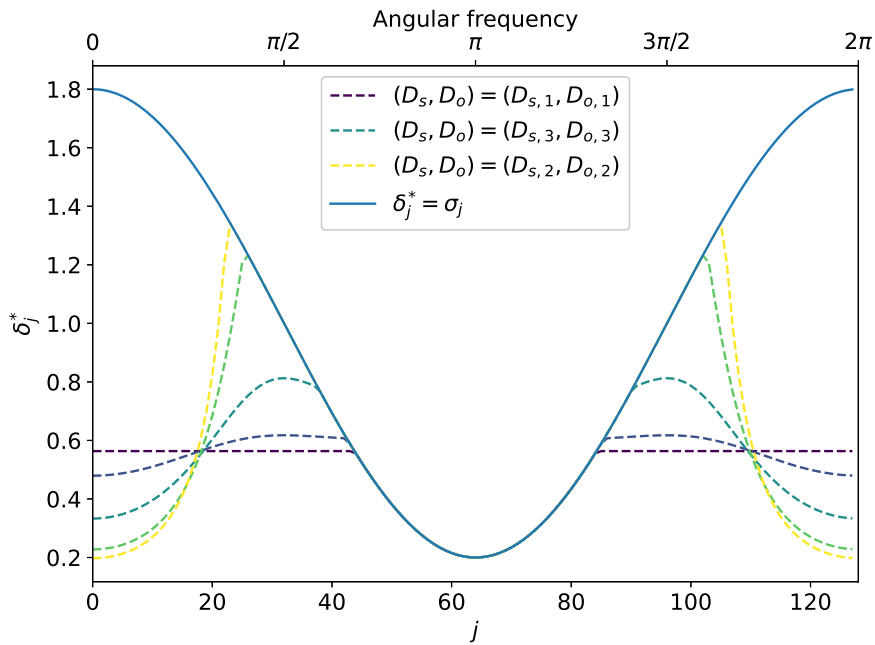}
    \caption{Optimal diagonal elements $( \delta_{1}^{*} , \delta_{2}^{*} , \delta_{3}^{*} )$ of $\mathbf{Q} \mathbf{\Delta} \mathbf{Q}^{T}$ for the marked points in Figure~\ref{f:water.cregions}, plotted with the colors in Figure~\ref{f:water.cregions}.}
    \label{f:water.cwater}
\end{figure}

In the illustrative example below, consider $\mathbf{K}_{X}$ as a $128 \times 128$ circulant matrix with the first row
\begin{equation*}
    [ 1 , 0.4 , 0 , \cdots , 0 , 0.4 ] ,
\end{equation*}
$\mathbf{H}$ as a $128 \times 128$ circulant matrix with the first row
\begin{equation*}
    [ 0.3, 0.3, 0.3, 0.3, 0 , \cdots , 0 ] ,
\end{equation*}
and $\mathbf{K}_{Z}$ as a $128 \times 128$ zero matrix (i.e., no noise in the state-observation relationship). Therefore, $\mathbf{Q}$ is the $128 \times 128$ DFT matrix whose $( i , j )$-th element is
\begin{equation*}
    \frac{1}{\sqrt{128}} e^{- \iu \frac{2 \pi}{128} i j}, \quad i, j = 0, 1, \cdots, 127.
\end{equation*}
The diagonal elements $\alpha_{0}$, $\alpha_{1}$, $\cdots$, $\alpha_{127}$ of $\mathbf{Q}^{\dag} \mathbf{H}^{T} \mathbf{H} \mathbf{Q}$ are shown in Figure~\ref{f:water.alpha}, and the diagonal elements $\sigma_{0}$, $\sigma_{1}$, $\cdots$, $\sigma_{127}$ of $\mathbf{Q}^{\dag} \mathbf{K}_{X} \mathbf{Q}$ are shown as the blue solid curve in Figure~\ref{f:water.cwater}. Figure~\ref{f:water.cregions} shows the four regions $A_{0}$, $A_{1}$, $A_{2}$, $A_{3}$ and the two curves $C_{s}$, $C_{o}$. It also displays five points on the contour of $R_\mathcal{G} ( D_{s} , D_{o} ) = 50$, marked with colors varying from purple to yellow. The weighted reverse water-filling solution $( \delta_{0}^{*} , \delta_{1}^{*} , \cdots , \delta_{127}^{*} )$ for these points are depicted in Figure~\ref{f:water.cwater}. For $( D_{s , 1} , D_{o , 1} )$, the optimal solution degenerates into a standard reverse water-filling form, as indicated by the purple line. When we go from $( D_{s , 1} , D_{o , 1} )$ to $( D_{s , 2} , D_{o , 2} )$, the water level begins to ``ripple''. Note that this weighted reverse water-filling can be viewed as exercised in the frequency domain, and the angular frequencies are marked on the top of Figure~\ref{f:water.cwater}.

\section{Conclusion}
\label{s:conclusion}

We have provided a general source model to describe information sources that have semantic aspects, and proposed a corresponding rate distortion problem formulation for characterizing the amount of information content of such semantic sources. We have studied the case of Gaussian extrinsic observation subject to a linear state-observation relationship and a quadratic distortion structure. There are a variety of issues that we have not touched upon in the present work. First, calculating and bounding the semantic rate distortion functions for other interesting cases would make further use of our proposed framework, for example, when the intrinsic state is a discrete categorical random variable, corresponding to the important problem of classification; see \cite{liu2021isit} for some preliminary results. Second, a more challenging problem is to estimate the semantic rate distortion function, and more importantly, to develop effective lossy compression methods when the joint probability distribution of the intrinsic state and the extrinsic observation is not perfectly known, say, when only finite training data of the state-observation pair are available.

\appendices

\section{Proof of Theorem \ref{thm:general.rdf}}
\label{ proof of theorem general case}

The key to proving Theorem \ref{thm:general.rdf} is converting the semantic rate distortion problem into an equivalent standard rate distortion problem, with an indirect (state) distortion constraint and a direct (observation) distortion constraint. More precisely, we need to show that the constraint with respect to the state distortion measure $d_s(s, \hat{s})$ is equivalent to a constraint on a converted distortion measure $\hat{d}_s(x, \hat{s})$; that is, as long as a reproduction $\hat{S}$ satisfies the constraint on $\hat{d}_s(x, \hat{s})$, it will satisfy the constraint on $d_s(s, \hat{s})$, and vice versa.

A general and unified approach to the indirect rate-distortion function put forward in \cite{witsenhausen1980} is first showing that the one-shot expected distortion $\mathbb{E} \left[ d_s(S, \hat{S}) \right]$ is equivalent to $\mathbb{E} \left[ \hat{d}_s (X, \hat{S}) \right]$, and then invoking a tensorization argument to extend the one-shot equivalence to block codes. Here we directly illustrate how this can be accomplished for $S^n \leftrightarrow X^n \leftrightarrow (\hat{S}^n, \hat{X}^n)$ generated by an arbitrary encoder-decoder pair, as follows:

\begin{eqnarray}
     \mathbb{E} \left[ d_s(S^n, \hat{S}^n) \right] &=& \sum_{s^n, \hat{s}^n} p(s^n, \hat{s}^n) d_s(s^n, \hat{s}^n)\nonumber\\
    &=& \sum_{s^n, x^n, \hat{s}^n} p(s^n, x^n, \hat{s}^n) d_s(s^n, \hat{s}^n)\nonumber\\
    &\stackrel{(a)}{=}& \sum_{s^n, x^n, \hat{s}^n} p(s^n,x^n) p(\hat{s}^n|x^n) d_s(s^n, \hat{s}^n)\nonumber\\
    &=& \sum_{x^n, \hat{s}^n} p(\hat{s}^n|x^n) \sum_{s^n} p(s^n ,x^n) d_s(s^n, \hat{s}^n)\nonumber\\
 &\stackrel{(b)}{=}& \sum_{x^n, \hat{s}^n} p(\hat{s}^n|x^n) \sum_{s^n} p(s^n ,x^n) \frac{1}{n}\sum_{i=1}^{n}d_s(s_i, \hat{s}_i)\nonumber\\
  &=& \sum_{x^n, \hat{s}^n} p(\hat{s}^n|x^n) \frac{1}{n}\sum_{i=1}^{n}\sum_{s^n} p(s^n ,x^n) d_s(s_i, \hat{s}_i)\nonumber\\
    &\stackrel{(c)}{=}& \sum_{x^n, \hat{s}^n} p(\hat{s}^n|x^n) \frac{1}{n}\sum_{i=1}^{n}\sum_{\bar{s}_i}\sum_{s_i\in \mathcal{S}}p(\bar{s}_i,\bar{x}_i) p(s_i ,x_i) d_s(s_i, \hat{s}_i)\nonumber\\
        &=& \sum_{x^n, \hat{s}^n} p(\hat{s}^n|x^n) \frac{1}{n}\sum_{i=1}^{n}\sum_{\bar{s}_i}p(\bar{s}_i,\bar{x}_i)\sum_{s_i\in \mathcal{S}} p(s_i ,x_i) d_s(s_i, \hat{s}_i)\nonumber\\
&=& \sum_{x^n, \hat{s}^n} p(\hat{s}^n|x^n) \frac{1}{n}\sum_{i=1}^{n}p(\bar{x}_i)\sum_{s_i\in \mathcal{S}} p(s_i ,x_i) d_s(s_i, \hat{s}_i)\nonumber\\
&=& \sum_{x^n, \hat{s}^n} p(\hat{s}^n|x^n) \frac{1}{n}\sum_{i=1}^{n}p(x^n)\sum_{s_i\in \mathcal{S}} p(s_i |x_i) d_s(s_i, \hat{s}_i)\nonumber\\
    &\stackrel{(d)}{=}& \sum_{x^n, \hat{s}^n} p(x^n, \hat{s}^n) \frac{1}{n} \sum_{i = 1}^n \hat{d}_s(x_i, \hat{s}_i)\nonumber\\
    &=& \sum_{x^n, \hat{s}^n} p(x^n, \hat{s}^n)  \hat{d}_s(x^n, \hat{s}^n)\nonumber\\
    &=& \mathbb{E} \left[ \hat{d}_s(X^n, \hat{S}^n) \right] , \label{e:proof1.indist} 
\end{eqnarray}
where $\bar{x}_i=(x_1,\dots,x_{i-1},x_{i+1},\dots,x_n)$, $\bar{s}_i=(s_1,\dots,s_{i-1},s_{i+1},\dots,s_n)$, $(a)$ is due to the existence of the Markov chain $S^n \leftrightarrow X^n \leftrightarrow \hat{S}^n$ and hence $p(s^n|x^n)=p(s^n|x^n,\hat{s}^n)$, $(b)$ follows from the definition of block-wise distortion measure in \eqref{block error def 1}, $(c)$ is by the fact that $(S_i,X_i)_{i\in \mathbf{N}}$ is an i.i.d. sequence, and $(d)$ is by the definition of $\hat{d}_s(x, \hat{s})$ in \eqref{e:general.reducedd}. 
Subsequently, the problem is reduced into a standard lossy source coding problem with two distortion constraints, one on $d_o(x, \hat{x})$ and the other on $\hat{d}_s(x, \hat{s})$. The semantic rate distortion function hence follows from standard achievability and converse proofs \cite[Sec. VII]{gamal1982} \cite[Prob. 7.14]{csiszar2011} \cite[Prob. 10.19]{cover2005}.

\section{Proof of Theorem~\ref{thm:gaussian.rdf}}
\label{u:gaussian.proof}

The proof of Theorem \ref{thm:gaussian.rdf} involves two steps. First we prove that the semantic rate distortion function can be achieved by jointly Gaussian $\hat{X}$ and $\hat{S}$. Then we show that we can further endow a Markov chain structure on $X$, $\hat{X}$ and $\hat{S}$, so that we only need to optimize with one variable, i.e., $\hat{X}$, while generating $\hat{S}$ from $\hat{X}$ subsequently.

\subsection{Optimality of Jointly Gaussian Reproduction}
\label{Proof of Gaussian Case}
By the definition of $\hat{d}_{s} ( X ; \hat{S} )$ in \eqref{e:general.reducedd},  $\mathbb{E} \left[ \hat{d}_{s} ( X ; \hat{S}) \right]$ can be written as follows:
\begin{align}
    \mathbb{E} \left[ \hat{d}_{s} ( X ; \hat{S}) \right] = &\int p(x,\hat{s})\left(\int p(s| x) d_s(s, \hat{s}) ds \right)dxd\hat{s}\nonumber\\
    = &\int p(x,\hat{s})\left(\int p(\mathbf{H}x+z| x) (\mathbf{H}x+z-\hat{s})(\mathbf{H}x+z-\hat{s})^T dz \right)dxd\hat{s}\nonumber\\
    \overset{(a)}{=} &\int p(x,\hat{s})\bigg(\int p(z) \tr (\mathbf{H}xx^T\mathbf{H}^T+\mathbf{H}xz^T-\mathbf{H}x\hat{s}\nonumber\\
    &+zx^T\mathbf{H}+zz^T-z\hat{s}^T-\hat{s}x^T\mathbf{H}^T-\hat{s}z^T+\hat{s}\hat{s}^T) dz\bigg)dxd\hat{s}\nonumber\\   
    \overset{(b)}{=}&\int p(x,\hat{s}) \tr (\mathbf{H}xx^T\mathbf{H}^T-\mathbf{H}x\hat{s} +\mathbf{K}_Z-\hat{s}x^T\mathbf{H}^T+\hat{s}\hat{s}^T) dxd\hat{s}\nonumber\\
    =&\tr(\mathbf{H}\mathbf{K}_X\mathbf{H}^T-\mathbf{H}\mathbf{K}_{X\hat{S}}+\mathbf{K}_Z-\mathbf{K}_{\hat{S}X}\mathbf{H}^T+\mathbf{K}_{\hat{S}})\nonumber\\
    \overset{(c)}{=}& \tr(\mathbf{H}\mathbf{K}_X\mathbf{H}^T-2\mathbf{H}\mathbf{K}_{X\hat{S}}+\mathbf{K}_Z+\mathbf{K}_{\hat{S}}), \label{e:proof2.ed}
\end{align}
where $(a)$ is due to independence between $Z$ and $X$, $(b)$ is according to the problem setup that $\mathbb{E}(Z)=0$, and $(c)$ is due to the fact that $\tr(\mathbf{H}\mathbf{K}_{X\hat{S}})=\tr(\mathbf{K}_{\hat{S}X}\mathbf{H}^T)$. From this chain of identities, we see that for any two reproductions of the intrinsic state, $\hat{S}$ and $\hat{S}'$, we have $\mathbb{E} \left[ d_{S} ( S ; \hat{S}) \right] = \mathbb{E} \left[ d_{S} ( S ; \hat{S}') \right]$ as long as $\mathbf{K}_{\hat{S}}=\mathbf{K}_{\hat{S}'}$ and $\mathbf{K}_{X\hat{S}}=\mathbf{K}_{X\hat{S}'}$.

Therefore, by Theorem~\ref{thm:general.rdf}, the semantic rate distortion function $R ( D_{s} , D_{o})$ can be further written as
\begin{align}
    R (D_{s},D_{o})&=\min I(X;\hat{S},\hat{X})=h(X)-\max h(X|\hat{S},\hat{X})\label{eq:extend1}\\
    \text{s.t.} \quad    & \tr(\mathbf{K}_X-2\mathbf{K}_{X\hat{X}}+\mathbf{K}_{\hat{X}})\leq D_{o}\label{eq:extend2}\\
    & \tr(\mathbf{H}\mathbf{K}_X\mathbf{H}^T-2\mathbf{H}\mathbf{K}_{X\hat{S}}+\mathbf{K}_Z+\mathbf{K}_{\hat{S}}) \leq D_{s}\label{eq:extend3}.
\end{align}
Notice that, by denoting $T \triangleq (\hat{S},\hat{X})$ for convenience, $h(X|\hat{S},\hat{X})$ can be upper bounded as 
\begin{align}
    & h(X|\hat{S},\hat{X})\nonumber\\
    ={}& h(X|T)\nonumber\\
    ={}& h(X-\mathbf{K}_{XT}\mathbf{K}^{-1}_{T}T|T)\nonumber\\
    \overset{(a)}{\leq}{} &h(X-\mathbf{K}_{XT}\mathbf{K}^{-1}_{T}T)\nonumber\\
     \overset{(b)}{\leq}{} &\frac{1}{2}\log \det(2\pi e \mathbf{K}_{X-\mathbf{K}_{XT}\mathbf{K}^{-1}_{T}T})\nonumber\\
     ={}& \frac{1}{2}\log \det(2\pi e \mathbf{K}_{X}-\mathbf{K}_{XT}\mathbf{K}^{-1}_{T}\mathbf{K}_{TX}), \label{eq:upper bound of condition entropy}
\end{align}
where $(a)$ is by the fact that conditioning reduces entropy, and equality holds when $X-\mathbf{K}_{XT}\mathbf{K}^{-1}_{T}T$ is independent of $T$; $(b)$ is due to the fact that Gaussian distribution maximizes differential entropy with given second central moment. Overall, we can see that this upper bound of $h(X|\hat{S},\hat{X})$ is achieved when $X$ and $T$ are jointly Gaussian.

Based on the argument above, for an arbitrary $T = (\hat{S}, \hat{X})$, we can generate $T' = (\hat{S}', \hat{X}')$ according to a linear relationship
\begin{align}
    (\hat{S}',\hat{X}')= \mathbf{K}_{TX}\mathbf{K}^{-1}_{X}X+N,
\end{align}
where $N$ is a multivariate Gaussian random variable following $\mathcal{N}(0,\mathbf{K}_{T}-\mathbf{K}_{TX}\mathbf{K}^{-1}_{X}\mathbf{K}_{XT})$ and is independent of $X$. Clearly it holds that $\mathbf{K}_{T'} = \mathbf{K}_T$ and $\mathbf{K}_{XT} = \mathbf{K}_{XT'}$. According to \eqref{eq:upper bound of condition entropy}, we can see that $h(X|\hat{S},\hat{X})\leq h(\hat{S}',\hat{X}')$. That is to say, for any $(\hat{S},\hat{X})$ that satisfies the distortion constraints, there always exists a Gaussian $(\hat{S}',\hat{X}')$ which also satisfies the distortion constraints, but achieving a lower code rate. We thus establish that jointly Gaussian reproduction $(\hat{S},\hat{X})$ achieves the semantic rate distortion function.

\subsection{Reduction to One Optimization Variable}
In fact, it is unnecessary to optimize with two random variables $(\hat{S},\hat{X})$ simultaneously,  
%
and in the following we reduce the number of optimization variables to only one. We choose the new optimization variable as $\mathrm{cov}(X|\hat{X},\hat{S})$, defined as
\begin{align*}
    \mathrm{cov}(X|\hat{X},\hat{S})=\mathbb{E}\left[\left(X-\mathbb{E}\left[X|\hat{X},\hat{S}\right]\right)\left(X-\mathbb{E}\left[X|\hat{X},\hat{S}\right]\right)^T\right],
\end{align*}
i.e., the error covariance matrix of MMSE estimating $X$ by $(\hat{X},\hat{S})$. 
By denoting $\mathrm{cov}(X|\hat{X},\hat{S})$ as $\mathbf{\Delta}$ for short, we can write $I(X;\hat{X},\hat{S})$ as \eqref{e:gaussian.rdf}. Therefore, now the key point is to show that the feasible region defined by \eqref{eq:extend2}-\eqref{eq:extend3} (denoted as $\mathcal{R}_1$) is the same as the feasible region defined by \eqref{e:gaussian.deltale}-\eqref{e:gaussian.trdx} (denoted as $\mathcal{R}_2$).

First we show that $\mathcal{R}_1 \subseteq \mathcal{R}_2$. For any $\mathbf{K}_{(\hat{S},\hat{X})} \in \mathcal{R}_1$, with $\mathbf{\Delta}=\mathrm{cov}(X|\hat{S},\hat{X})$, we have $\mathbf{\Delta}\preceq \mathrm{cov}(X|\hat{X})$ and $\mathbf{\Delta} \preceq \mathrm{cov}(X|\hat{S})$,
and correspondingly $\tr(\mathbf{\Delta}) \leq \tr(\mathrm{cov}(X|\hat{X})) \leq D_{o}$ and
\begin{align}
    \tr(\mathbf{H}\mathbf{\Delta} \mathbf{H}^T+\mathbf{K}_Z)
    \leq \tr(\mathbf{H} \mathrm{cov}(X|\hat{S})\mathbf{H}^T+\mathbf{K}_Z)=\tr(\mathrm{cov}(\mathbf{H}X+Z|\hat{S})) \leq D_{s}.
\end{align}
That is to say, for any $\mathbf{K}_{(\hat{S},\hat{X})} \in \mathcal{R}_1$, we can find a corresponding $\mathbf{\Delta} \in \mathcal{R}_2$, and hence $\mathcal{R}_1 \subseteq \mathcal{R}_2$.

Then we show that $\mathcal{R}_2 \subseteq \mathcal{R}_1$. For any $\mathbf{\Delta} \in \mathcal{R}_2$, we consider a test channel with $X = \hat{X} + N$ and let $\hat{S}=\mathbf{H}\hat{X}$, where $N$ obeys Gaussian distribution $\mathcal{N}(0, \mathbf{\Delta})$. Hence we have
\begin{align}
    & \mathbb{E} \left[d_{o}(X,\hat{X})\right] =\tr(\mathbf{\Delta}) \leq D_o,\\
    & \mathbb{E} \left[d_{s}(S,\hat{S})\right] =\tr(\mathbf{H} \mathrm{cov}(X|\hat{S})\mathbf{H}^T+\mathbf{K}_Z)=\tr(\mathbf{H}\mathbf{\Delta} \mathbf{H}^T+\mathbf{K}_Z)\leq D_s.
\end{align}
That is to say, for any $\mathbf{\Delta} \in \mathcal{R}_2$, we can also find a corresponding tuple of $\mathbf{K}_{(\hat{S},\hat{X})} \in \mathcal{R}_1$, and hence $\mathcal{R}_2 \subseteq \mathcal{R}_1$.

Now, we can conclude that, under the setting of Theorem \ref{thm:gaussian.rdf}, Theorems \ref{thm:general.rdf} and \ref{thm:gaussian.rdf} define two optimization problems with the same objective function and the same feasible region. This therefore completes the proof.

\section{Proof of Corollary~\ref{cor:gaussian.largest_rdf}}
\label{s:largest_rdf}


By Theorem~\ref{thm:gaussian.rdf} and the identities $\mathbf{H} = \mathbf{K}_{S X} \mathbf{K}_{X}^{- 1}$ and $\mathbf{K}_{Z} = \mathbf{K}_{S} - \mathbf{K}_{S X} \mathbf{K}_{X}^{- 1} \mathbf{K}_{S X}^{T}$ in \eqref{e:gaussian.linear}, the semantic rate distortion function of a jointly Gaussian semantic source with covariance matrix \eqref{e:gaussian.covmat} is given by
\begin{align}
    R_{\mathcal{G}} ( D_{s} , D_{o} )
    & = \min_{\mathbf{\Delta} \in \mathcal{S}_m}  \frac{1}{2}
    \log \left( \frac{\det ( \mathbf{K}_{X} )}{\det ( \mathbf{\Delta} )} \right) \label{e:largest_rdf.bound} \\
    \text{s.t.} \quad  & \mathbf{O} \prec \mathbf{\Delta} \preceq \mathbf{K}_{X} ,
    \label{e:largest_rdf.deltale} \\
    & \tr (
        \mathbf{K}_{S X} \mathbf{K}_{X}^{- 1} \mathbf{\Delta}
        \mathbf{K}_{X}^{- 1} \mathbf{K}_{S X}^{T}
    )
    \le D_{s} - \tr (
        \mathbf{K}_{S}
        - \mathbf{K}_{S X} \mathbf{K}_{X}^{- 1} \mathbf{K}_{S X}^{T}
    ) ,
    \label{e:largest_rdf.trds} \\
    & \tr ( \mathbf{\Delta} ) \le D_{o} .
    \label{e:largest_rdf.trdo}
\end{align}
We will prove
\begin{equation}
    R ( D_{s} , D_{o} )
    \le \frac{1}{2}
    \log \left( \frac{\det ( \mathbf{K}_{X} )}{\det ( \mathbf{\Delta} )} \right)
    \label{e:largest_rdf.ratele}
\end{equation}
for an arbitrary symmetric matrix $\mathbf{\Delta}$ that satisfies \eqref{e:largest_rdf.deltale}, \eqref{e:largest_rdf.trds} and \eqref{e:largest_rdf.trdo}, by constructing a test channel. This implies that $R ( D_{s} , D_{o} )$ is no greater than \eqref{e:largest_rdf.bound}.

In order to construct the test channel, let $U$ be a Gaussian vector with zero mean and covariance matrix $\mathbf{\Delta} - \mathbf{\Delta} \mathbf{K}_{X}^{- 1} \mathbf{\Delta}$, independent of $( S , X )$. That $\mathbf{\Delta} - \mathbf{\Delta} \mathbf{K}_{X}^{- 1} \mathbf{\Delta}$ is semi-definite will be proved in 
Lemma~\ref{lem:largest_rdf.psd} at the end of this subsection. Define $\hat{X} = ( \mathbf{I}_{m} - \mathbf{\Delta} \mathbf{K}_{X}^{- 1} ) X + U$ and $\hat{S} = \mathbf{K}_{S X} \mathbf{K}_{X}^{- 1} \hat{X}$. Thus $S \leftrightarrow X \leftrightarrow \hat{X} \leftrightarrow \hat{S}$ is a Markov chain. We will verify in the next paragraphs that $\mathbb{E} [ \hat{d}_{s} ( X , \hat{S} ) ] \le D_{s}$, where $\hat{d}_{s} ( x , \hat{s} ) = \mathbb{E} [ \| S - \hat{s} \|_{2}^{2} | X = x ]$, $\mathbb{E} [ \| X - \hat{X} \|_{2}^{2} ] \le D_{o}$, and
\begin{equation}
    I ( X ; \hat{S} , \hat{X} )
    \le \frac{1}{2}
    \log \left( \frac{\det ( \mathbf{K}_{X} )}{\det ( \mathbf{\Delta} )} \right) .
    \label{e:largest_rdf.minfole} 
\end{equation}
These leads to \eqref{e:largest_rdf.ratele}, and thus proves Corollary~\ref{cor:gaussian.largest_rdf}.

By the definitions of $\hat{X}$ and $\hat{S}$, we have $S - \hat{S} = S - \mathbf{K}_{S X} \mathbf{L} X - \mathbf{K}_{S X} \mathbf{K}_{X}^{- 1} U$, where $\mathbf{L} = \mathbf{K}_{X}^{- 1} - \mathbf{K}_{X}^{- 1} \mathbf{\Delta} \mathbf{K}_{X}^{- 1}$. Noticing $\mathbb{E} [ S U^{T} ] = \mathbf{O}_{l \times m}$ and $\mathbb{E} [ X U^{T} ] = \mathbf{O}_{m \times m}$, we can obtain, after some algebraic manipulations,
\begin{equation*}
    \mathbb{E} \left[ ( S - \hat{S} ) ( S - \hat{S} )^{T} \right]
    = \mathbf{K}_{S}
    - \mathbf{K}_{S X} \mathbf{K}_{X}^{- 1} \mathbf{K}_{S X}^{T}
    + \mathbf{K}_{S X} \mathbf{K}_{X}^{- 1} \mathbf{\Delta}
    \mathbf{K}_{X}^{- 1} \mathbf{K}_{S X}^{T}.
\end{equation*}
Taking the trace in this equation and using \eqref{e:largest_rdf.trds}, we get $\mathbb{E} [ \| S - \hat{S} \|_{2}^{2} ] \le D_{s}$. Similar calculations lead to $\mathbb{E} [ \| X - \hat{X} \|_{2}^{2} ] \le D_{o}$. For every $x \in \reals^{m}$ and every $\hat{s} \in \reals^{l}$, we have
\begin{equation*}
    \mathbb{E} [ \| S - \hat{S} \|_{2}^{2} | X = x , \hat{S} = \hat{s} ]
    = \mathbb{E} [ \| S - \hat{s} \|_{2}^{2} | X = x , \hat{S} = \hat{s} ]
    = \mathbb{E} [ \| S - \hat{s} \|_{2}^{2} | X = x ]
    = \hat{d}_{s} ( x , \hat{s} ) ,
\end{equation*}
where the second equality is due to $S \leftrightarrow X \leftrightarrow \hat{S}$. An application of the law of total expectation immediately leads to $\mathbb{E} [ \hat{d}_{s} ( X , \hat{S} ) ] = \mathbb{E} [ \| S - \hat{S} \|_{2}^{2} ] \le D_{s}$.


It remains to verify \eqref{e:largest_rdf.minfole}. We have
\begin{align*}
    I ( X ; \hat{S} , \hat{X} )
    & \stackrel{\text{(a)}}{=} I ( X ; \hat{X} ) \\
    & = h ( \hat{X} ) - h ( \hat{X} | X ) \\
    & \stackrel{\text{(b)}}{=} h ( \hat{X} )
    - \frac{1}{2} \log ( ( 2 \pi e )^{m} \det (
        \mathbf{\Delta}
        - \mathbf{\Delta} \mathbf{K}_{X}^{- 1} \mathbf{\Delta}
    ) ) \\
    & \stackrel{\text{(c)}}{\le} \frac{1}{2}
    \log ( ( 2 \pi e )^{m} ( \mathbf{K}_{X} - \mathbf{\Delta} ) )
    - \frac{1}{2} \log ( ( 2 \pi e )^{m} \det (
        \mathbf{\Delta}
        - \mathbf{\Delta} \mathbf{K}_{X}^{- 1} \mathbf{\Delta}
    ) ) \\
    & = \frac{1}{2}
    \log \left( \frac{\det ( \mathbf{K}_{X} )}{\det ( \mathbf{\Delta} )} \right) ,
\end{align*}
where (a) is by $X \leftrightarrow \hat{X} \leftrightarrow \hat{S}$, (b) is because after translation $h(\hat{X}|X) = h(U)$, and (c) is because the Gaussian distribution maximizes the differential entropy subject to a covariance constraint.

Finally let us verify the existence of the auxiliary random vector $U$.
\begin{lemma}
    \label{lem:largest_rdf.psd} 
For any $\mathbf{\Delta}$, $\mathbf{K} \in \mathcal{S}_{m}$, $\mathbf{\Delta} \preceq \mathbf{K}$, $\mathbf{\Delta} - \mathbf{\Delta} \mathbf{K}^{- 1} \mathbf{\Delta}$ is semi-definite.
\end{lemma}

\textit{Proof:} Because $\mathbf{\Delta}$ is positive definite, there exists an $m \times m$ matrix $\mathbf{Q}$ such that $\mathbf{\Delta} = \mathbf{Q}^{T} \mathbf{Q}$. For every $\lambda < 0$,
\begin{align*}
    \det (
        \lambda \mathbf{I}_{m}
        - ( \mathbf{I}_{m} - \mathbf{Q} \mathbf{K}^{- 1} \mathbf{Q}^{T} )
    )
    & = ( \lambda - 1 )^{m} \det \left(
        \mathbf{I}_{m}
        + \frac{1}{\lambda - 1} \mathbf{Q} \mathbf{K}^{- 1} \mathbf{Q}^{T}
    \right) \\
    & = ( \lambda - 1 )^{m} \det \left(
        \mathbf{I}_{m}
        + \frac{1}{\lambda - 1} \mathbf{K}^{- 1} \mathbf{Q}^{T} \mathbf{Q}
    \right) \\
    & = ( \lambda - 1 )^{m} \det \left(
        \mathbf{K}^{- 1}
        \left( \mathbf{K} + \frac{1}{\lambda - 1} \mathbf{\Delta} \right)
    \right) \\
    & = \frac{
        \det ( ( \lambda - 1 ) \mathbf{K} + \mathbf{\Delta} )
    }{\det ( \mathbf{K} )} \not= 0 ,
\end{align*}
because $( \lambda - 1 ) \mathbf{K} + \mathbf{\Delta} = \lambda \mathbf{K} - ( \mathbf{K} - \mathbf{\Delta} )$ is negative definite. So $\mathbf{I}_{m} - \mathbf{Q} \mathbf{K}^{- 1} \mathbf{Q}^{T}$ does not have any negative eigenvalue. Therefore $\mathbf{I}_{m} - \mathbf{Q} \mathbf{K}^{- 1} \mathbf{Q}^{T}$ is positive semi-definite, and consequently $\mathbf{\Delta} - \mathbf{\Delta} \mathbf{K}^{- 1} \mathbf{\Delta} = \mathbf{Q}^{T} ( \mathbf{I}_{m} - \mathbf{Q} \mathbf{K}^{- 1} \mathbf{Q}^{T} ) \mathbf{Q}$ is also positive semi-definite. $\Box$

\section{Derivation of the Weighted Reverse Water-Filling Solution}
\label{s:waterderi}

We first rewrite \eqref{e:gaussian.rdf} with a variable substitution $\mathbf{\Delta} = \mathbf{Q} \mathbf{D} \mathbf{Q}^{\dag}$. This leads to
\begin{equation*}
    R_\mathcal{G} ( D_{s} , D_{o} )
    = \min_{\mathbf{D} \in B ( D_{s} , D_{o} )}
    \frac{1}{2}
    \log \left( \frac{\det ( \mathbf{K}_{X} )}{\det ( \mathbf{D} )} \right) ,
\end{equation*}
where $B ( D_{s} , D_{o} )$ is the set of positive definite real matrices $\mathbf{D}$ that satisfy
\begin{align*}
    \mathbf{D} & \preceq \mathbf{Q}^{\dag} \mathbf{K}_{X} \mathbf{Q} , \\
    \tr ( \mathbf{Q}^{\dag} \mathbf{H}^{T} \mathbf{H} \mathbf{Q} \mathbf{D} ) & \le D_{s} - \tr ( \mathbf{K}_{Z} ) , \\
    \tr ( \mathbf{D} ) & \le D_{o} .
\end{align*}
Any optimal $\mathbf{D}$ in this minimization is diagonal. To see this, consider a non-diagonal $\mathbf{D} \in B ( D_{s} , D_{o} )$. Replacing the non-diagonal elements in $\mathbf{D}$ with zeros, we get a new matrix $\mathbf{D}' = \diag ( \delta_{1} , \delta_{2} , \cdots , \delta_{m} )$. Because
\begin{equation*}
    \mathbf{O}_{m}
    \prec \mathbf{D}
    \preceq \mathbf{Q}^{\dag} \mathbf{K}_{X} \mathbf{Q}
    = \diag ( \sigma_{1} , \sigma_{2} , \cdots , \sigma_{m} ) ,
\end{equation*}
we have $0 < \delta_{j} \le \sigma_{j}$ for each $j \in \{ 1 , 2 , \cdots , m \}$, which impies $\mathbf{O}_{m} \prec \mathbf{D}' \preceq \mathbf{Q}^{\dag} \mathbf{K}_{X} \mathbf{Q}$. Moreover,
\begin{align*}
    \tr ( \mathbf{Q}^{\dag} \mathbf{H}^{T} \mathbf{H} \mathbf{Q} \mathbf{D}' )
    & = \sum_{j = 1}^{m} \alpha_{j} \delta_{j}
    = \tr ( \mathbf{Q}^{\dag} \mathbf{H}^{T} \mathbf{H} \mathbf{Q} \mathbf{D} )
    \le D_{s} - \tr ( \mathbf{K}_{Z} ) ,
\\
    \tr ( \mathbf{D}' )
    & = \sum_{j = 1}^{m} \delta_{j}
    = \tr ( \mathbf{D} )
    \le D_{o} .
\end{align*}
So $\mathbf{D}' \in B ( D_{s} , D_{o} )$. By Hadamard's inequality,
\begin{equation*}
    \frac{1}{2}
    \log \left( \frac{\det ( \mathbf{K}_{X} )}{\det ( \mathbf{D}' )} \right)
    < \frac{1}{2}
    \log \left( \frac{\det ( \mathbf{K}_{X} )}{\det ( \mathbf{D} )} \right) .
\end{equation*}
Therefore, any non-diagonal $\mathbf{D} \in B ( D_{s} , D_{o} )$ is suboptimal, and \eqref{e:water.rate} is verified.

By the Karush-Kuhn-Tucker (KKT) optimality conditions, there exist non-negative numbers $\lambda$, $\mu$, $\nu_{1}$, $\nu_{2}$, $\cdots$, $\nu_{m}$ that satisfy
\begin{align*}
    \lambda
    \left( \sum_{j = 1}^{m} \delta_{j}^{*} - D_{o} \right)
    = 0 , &
\\
    \mu \left(
        \sum_{j = 1}^{m} \alpha_{j} \delta_{j}^{*}
        - D_{s}
        + \tr ( \mathbf{K}_{Z} )
    \right)
    = 0 , &
\\
    \nu_{j} ( \delta_{j}^{*} - \sigma_{j} ) = 0 , &
    \quad \forall j \in \{ 1 , 2 , \cdots , m \} ,
\\
    - \frac{1}{\delta_{j}^{*}} + \lambda + \mu \alpha_{j} + \nu_{j}
    = 0 , &
    \quad \forall j \in \{ 1 , 2 , \cdots , m \} .
\end{align*}

Suppose $\lambda = 0$ and $\mu = 0$. For each $j \in \{ 1 , 2 , \cdots , m \}$, we have $\nu_{j} = 1 / \delta_{j}^{*} > 0$, so $\delta_{j}^{*} = \sigma_{j}$. Because $\delta_{1}^{*}$, $\delta_{2}^{*}$, $\cdots$, $\delta_{m}^{*}$ satisfy \eqref{e:water.sumds} and \eqref{e:water.sumdx}, we have
\begin{align*}
    D_{s}
    & \ge \sum_{j = 1}^{m} \alpha_{j} \sigma_{j}
    + \tr ( \mathbf{K}_{Z} )
    = \tr ( \mathbf{H} \mathbf{K}_{X} \mathbf{H}^{T} + \mathbf{K}_{Z} ) , \\
    D_{o}
    & \ge \sum_{j = 1}^{m} \sigma_{j}
    = \tr ( \mathbf{K}_{X} ) ,
\end{align*}
i.e. $( D_{s} , D_{o} ) \in A_{0}$.

Suppose $\lambda > 0$ and $\mu = 0$. The problem now reduces to the one involved in the rate distortion problem of parallel Gaussian sources \cite{cover2005}, because the constraint \eqref{e:water.sumdx} is active and \eqref{e:water.sumds} is not. Thus \eqref{e:water.solution1} holds, and
\begin{align*}
    D_{o}
    & = \sum_{j = 1}^{m} \delta_{j}^{*}
    = \sum_{j = 1}^{m}
    \min \left( \sigma_{j} , \frac{1}{\lambda} \right) ,
\\
    D_{s}
    & \ge \sum_{j = 1}^{m} \alpha_{j} \delta_{j}^{*}
    + \tr ( \mathbf{K}_{Z} )
    = \sum_{j = 1}^{m}
    \alpha_{j}
    \min \left( \sigma_{j} , \frac{1}{\lambda} \right)
    + \tr ( \mathbf{K}_{Z} ) ,
\end{align*}
i.e. $( D_{s} , D_{o} ) \in A_{1}$.

Similarly, the conditions $\lambda = 0$ and $\mu > 0$ imply \eqref{e:water.solution2} leading to $( D_{s} , D_{o} ) \in A_{2}$, and the conditions $\lambda > 0$ and $\mu > 0$ imply \eqref{e:water.solution3} leading to $( D_{s} , D_{o} ) \in A_{3}$.

\bibliographystyle{IEEEtran}
\bibliography{IEEEabrv,bibliography}

\end{document}